 
 \def\figure#1#2{\centerline{\bf Fig. #1:}\par\noindent{\fig #2}} 
 \catcode`\"=\active\let"=\"


 \def\of{\overline{f}}

 \def\3{\ss }
 
 \newcount\notenumber  \notenumber=0
 \def\note{\advance\notenumber by1 \footnote{$^{\the\notenumber}$}}

 \def\parder#1#2{{\partial #1\over\partial #2}}
 \def\div#1{{1\over r^2}\parder{}{r}\left(r^2 #1 \right)}

 \def\Eq#1{(\the\ueberzaehla .#1)}
 \def\Fig#1{\the\ueberzaehla .#1}

 \def\simgr{\,\raise 0.3ex\hbox{$>$}\kern -0.8em\lower 0.7ex
                 \hbox{$\sim $}\,}
 \def\simlt{\,\raise 0.3ex\hbox{$<$}\kern -0.8em\lower 0.7ex
                 \hbox{$\sim $}\,}
   \def\ApJ{Astrophys. J.}
   
   \def\AA{Astron. Astrophys.}
   
   \def\MN{Monthly Notices Roy. Astron. Soc.}
   
   \def\Nat{Nature}

   \def\SA{Soviet Astron.}

   \def\PASJ{Publ. astr. Soc. Jap.}

  %
  
 \def\usrref#1#2#3#4#5{\ref #2: #3, {\it #1}~{\bf #4}, #5}
 
 \def\figure#1#2{\centerline{\bf Fig. #1:}\par\noindent{\fig #2}} 
 \catcode`\"=\active\let"=\"

 \def\of{\overline{f}}

 \def\3{\ss }
 
 \newcount\notenumber  \notenumber=0
 \def\note{\advance\notenumber by1 \footnote{$^{\the\notenumber}$}}

 \def\parder#1#2{{\partial #1\over\partial #2}}
 \def\div#1{{1\over r^2}\parder{}{r}\left(r^2 #1 \right)}
 \def\div4#1{{1\over r^4}\parder{}{r}\left(r^4 #1 \right)}

 \def\Eq#1{(\the\ueberzaehla .#1)}
 \def\Fig#1{\the\ueberzaehla .#1}

 \def\simgr{\,\raise 0.3ex\hbox{$>$}\kern -0.8em\lower 0.7ex
                 \hbox{$\sim $}\,}
 \def\simlt{\,\raise 0.3ex\hbox{$<$}\kern -0.8em\lower 0.7ex
                 \hbox{$\sim $}\,}
   \def\ApJ{Astrophys. J.}
   
   \def\AA{Astron. Astrophys.}
   
   \def\MN{Monthly Notices Roy. Astron. Soc.}
   
   \def\Nat{Nature}

   \def\SA{Soviet Astron.}

   \def\PASJ{Publ. astr. Soc. Jap.}

   \def\NewA{New Astronomy} 
  %
  
 \def\usrref#1#2#3#4#5{\ref #2: #3, {\it #1}~{\bf #4}, #5}

\overfullrule=0pt
\tolerance=6000
\font\SCS=cmcsc10 scaled 1000

%
\def\SpringerMacroPackageNameATest{AA}%
\let\next\relax
\ifx\SpringerMacroPackageNameA\undefined
  \message{Loading the \SpringerMacroPackageNameATest\space
           macro package from Springer-Verlag...}%
\else
  \ifx\SpringerMacroPackageNameA\SpringerMacroPackageNameATest
    \message{\SpringerMacroPackageNameA\space macro package
             from Springer-Verlag already loaded.}%
    \let\next\endinput
  \else
    \message{DANGER: \SpringerMacroPackageNameA\space from
             Springer-Verlag already loaded, will try to proceed.}%
  \fi
\fi
\next
\def\SpringerMacroPackageNameA{AA}%
\newskip\mathindent      \mathindent=0pt
\newskip\tabefore \tabefore=20dd plus 10pt minus 5pt      
\newskip\taafter  \taafter=10dd                           
\newskip\tbbeforeback    \tbbeforeback=-20dd              
\newskip\tbbefore        \tbbefore=17pt plus 7pt minus3pt 
\newskip\tbafter         \tbafter=8pt                     
\newskip\tcbeforeback    \tcbeforeback=-3pt               
\advance\tcbeforeback by -10dd                            
\newskip\tcbefore        \tcbefore=10dd plus 5pt minus 1pt
\newskip\tcafter         \tcafter=6pt                     
\newskip\tdbeforeback    \tdbeforeback=-3pt                  
\advance\tdbeforeback by -10dd                               
\newskip\tdbefore        \tdbefore=10dd plus 4pt minus 1pt   
\newskip\petitsurround
\petitsurround=6pt\relax
\newskip\ackbefore      \ackbefore=10dd plus 5pt             
\newskip\ackafter       \ackafter=6pt                        
\newdimen\itemindent    \newdimen\itemitemindent
\itemindent=1.5em       \itemitemindent=2\itemindent
 \font \tatt            = cmbx10 scaled \magstep3
 \font \tats            = cmbx10 scaled \magstep1
 \font \tamt            = cmmib10 scaled \magstep3
 \font \tams            = cmmib10 scaled \magstep1
 \font \tamss           = cmmib10
 \font \tast            = cmsy10 scaled \magstep3
 \font \tass            = cmsy10 scaled \magstep1
 \font \tbtt            = cmbx10 scaled \magstep2
 \font \tbmt            = cmmib10 scaled \magstep2
 \font \tbst            = cmsy10 scaled \magstep2
\catcode`@=11    
\vsize=23.5truecm
\hoffset=-1true cm
\voffset=-1true cm
\normallineskip=1dd
\normallineskiplimit=0dd
\newskip\ttglue%
\def\ifundefin@d#1#2{%
\expandafter\ifx\csname#1#2\endcsname\relax}
\def\getf@nt#1#2#3#4{%
\ifundefin@d{#1}{#2}%
\global\expandafter\font\csname#1#2\endcsname=#3#4%
\fi\relax
}
\newfam\sffam
\newfam\scfam
\def\makesize#1#2#3#4#5#6#7{%
 \getf@nt{rm}{#1}{cmr}{#2}%
 \getf@nt{rm}{#3}{cmr}{#4}%
 \getf@nt{rm}{#5}{cmr}{#6}%
 \getf@nt{mi}{#1}{cmmi}{#2}%
 \getf@nt{mi}{#3}{cmmi}{#4}%
 \getf@nt{mi}{#5}{cmmi}{#6}%
 \getf@nt{sy}{#1}{cmsy}{#2}%
 \getf@nt{sy}{#3}{cmsy}{#4}%
 \getf@nt{sy}{#5}{cmsy}{#6}%
 \skewchar\csname mi#1\endcsname ='177
 \skewchar\csname mi#3\endcsname ='177
 \skewchar\csname mi#5\endcsname ='177
 \skewchar\csname sy#1\endcsname ='60
 \skewchar\csname sy#3\endcsname='60
 \skewchar\csname sy#5\endcsname='60
\expandafter\def\csname#1size\endcsname{%
 \normalbaselineskip=#7
 \normalbaselines
 \setbox\strutbox=\hbox{\vrule height0.75\normalbaselineskip%
    depth0.25\normalbaselineskip width0pt}%
 \textfont0=\csname rm#1\endcsname
 \scriptfont0=\csname rm#3\endcsname
 \scriptscriptfont0=\csname rm#5\endcsname
    \def\oldstyle{\fam1\csname mi#1\endcsname}%
 \textfont1=\csname mi#1\endcsname
 \scriptfont1=\csname mi#3\endcsname
 \scriptscriptfont1=\csname mi#5\endcsname
 \textfont2=\csname sy#1\endcsname
 \scriptfont2=\csname sy#3\endcsname
 \scriptscriptfont2=\csname sy#5\endcsname
 \textfont3=\tenex\scriptfont3=\tenex\scriptscriptfont3=\tenex
   \def\rm{%
 \fam0\csname rm#1\endcsname%
   }%
   \def\it{%
 \getf@nt{it}{#1}{cmti}{#2}%
 \textfont\itfam=\csname it#1\endcsname
 \fam\itfam\csname it#1\endcsname
   }%
   \def\sl{%
 \getf@nt{sl}{#1}{cmsl}{#2}%
 \textfont\slfam=\csname sl#1\endcsname
 \fam\slfam\csname sl#1\endcsname}%
   \def\bf{%
 \getf@nt{bf}{#1}{cmbx}{#2}%
 \getf@nt{bf}{#3}{cmbx}{#4}%
 \getf@nt{bf}{#5}{cmbx}{#6}%
 \textfont\bffam=\csname bf#1\endcsname
 \scriptfont\bffam=\csname bf#3\endcsname
 \scriptscriptfont\bffam=\csname bf#5\endcsname
 \fam\bffam\csname bf#1\endcsname}%
   \def\tt{%
 \getf@nt{tt}{#1}{cmtt}{#2}%
 \textfont\ttfam=\csname tt#1\endcsname
 \fam\ttfam\csname tt#1\endcsname
 \ttglue=.5em plus.25em minus.15em
   }%
  \def\sf{%
\getf@nt{sf}{#1}{cmss}{10 at #2pt}%
\textfont\sffam=\csname sf#1\endcsname
\fam\sffam\csname sf#1\endcsname}%
   \def\sc{%
 \getf@nt{sc}{#1}{cmcsc}{10 at #2pt}%
 \textfont\scfam=\csname sc#1\endcsname
 \fam\scfam\csname sc#1\endcsname}%
\rm }}
\makesize{IXf}{9}{VIf}{6}{Vf}{5}{10.00dd}
\def\normalsize{\IXfsize
\def\sf{%
   \getf@nt{sf}{IXf}{cmss}{9}%
   \getf@nt{sf}{VIf}{cmss}{10 at 6pt}%
   \getf@nt{sf}{Vf}{cmss}{10 at 5pt}%
   \textfont\sffam=\csname sfIXf\endcsname
   \scriptfont\sffam=\csname sfVIf\endcsname
   \scriptscriptfont\sffam=\csname sfVf\endcsname
   \fam\sffam\csname sfIXf\endcsname}%
}%
\newfam\mibfam
\def\mib{%
   \getf@nt{mib}{IXf}{cmmib}{10 at9pt}%
   \getf@nt{mib}{VIf}{cmmib}{10 at6pt}%
   \getf@nt{mib}{Vf}{cmmib}{10 at5pt}%
   \textfont\mibfam=\csname mibIXf\endcsname
   \scriptfont\mibfam=\csname mibVIf\endcsname
   \scriptscriptfont\mibfam=\csname mibVf\endcsname
   \fam\mibfam\csname mibIXf\endcsname}%
\makesize{Xf}{10}{VIf}{6}{Vf}{5}{10.00dd}
\Xfsize
\it\bf\tt\rm

\def\tentt{\ttXf}

\normalsize
\it\bf\tt\sf\mib\rm
\def\boldmath{\textfont1=\mibIXf \scriptfont1=\mibVIf
\scriptscriptfont1=\mibVf}
\newdimen\fullhsize
\newcount\verybad \verybad=1010
\let\lr=L%
\fullhsize=40cc
\hsize=19.5cc
\def\fullline{\hbox to\fullhsize}
\def\makefootline{\baselineskip=10dd \fullline{\the\footline}}
\def\makeheadline{\vbox to 0pt{\vskip-22.5pt
            \fullline{\vbox to 8.5pt{}\the\headline}\vss}\nointerlineskip}
\hfuzz=2pt
\vfuzz=2pt
\tolerance=1000
\abovedisplayskip=3 mm plus6pt minus 4pt
\belowdisplayskip=3 mm plus6pt minus 4pt
\abovedisplayshortskip=0mm plus6pt
\belowdisplayshortskip=2 mm plus4pt minus 4pt
\parindent=1.5em
\newdimen\stdparindent\stdparindent\parindent
\frenchspacing
\nopagenumbers
\predisplaypenalty=600        
\displaywidowpenalty=2000     
\def\widowsandclubs#1{\global\verybad=#1
   \global\widowpenalty=\the\verybad1      
   \global\clubpenalty=\the\verybad2  }    
\widowsandclubs{1010}
\def\paglay{\headline={{\normalsize\hsize=.75\fullhsize\ifnum\pageno=1
\vbox{\hrule\line{\vrule\kern3pt\vbox{\kern3pt
\hbox{\bf A\&A manuscript no.}
\hbox{(will be inserted by hand later)}
\kern3pt\hrule\kern3pt
\hbox{\bf Your thesaurus codes are:}
\hbox{\rightskip=0pt plus3em\advance\hsize by-7pt
\vbox{\bf\noindent\ignorespaces\the\THESAURUS}}
\kern3pt}\hfil\kern3pt\vrule}\hrule}
\rlap{\quad\AALogo}\hfil
\else\normalsize\ifodd\pageno\hfil\folio\else\folio\hfil\fi\fi}}}
\makesize{VIIIf}{8}{VIf}{6}{Vf}{5}{9.00dd}
      \getf@nt{sf}{VIIIf}{cmss}{8}%
      \getf@nt{sf}{VIf}{cmss}{10 at 6pt}%
      \getf@nt{sf}{Vf}{cmss}{10 at 5pt}%
      \getf@nt{mib}{VIIIf}{cmmib}{10 at 8pt}%
      \getf@nt{mib}{VIf}{cmmib}{10 at 6pt}%
      \getf@nt{mib}{Vf}{cmmib}{10 at 5pt}%
\VIIIfsize\it\bf\tt\rm
\normalsize
\def\petit{\VIIIfsize
   \def\sf{%
      \getf@nt{sf}{VIIIf}{cmss}{8}%
      \getf@nt{sf}{VIf}{cmss}{10 at 6pt}%
      \getf@nt{sf}{Vf}{cmss}{10 at 5pt}%
      \textfont\sffam=\csname sfVIIIf\endcsname
      \scriptfont\sffam=\csname sfVIf\endcsname
      \scriptscriptfont\sffam=\csname sfVf\endcsname
      \fam\sffam\csname sfVIIIf\endcsname
}%
\def\mib{%
   \getf@nt{mib}{VIIIf}{cmmib}{10 at 8pt}%
   \getf@nt{mib}{VIf}{cmmib}{10 at 6pt}%
   \getf@nt{mib}{Vf}{cmmib}{10 at 5pt}%
   \textfont\mibfam=\csname mibVIIIf\endcsname
   \scriptfont\mibfam=\csname mibVIf\endcsname
   \scriptscriptfont\mibfam=\csname mibVf\endcsname
   \fam\mibfam\csname mibIXf\endcsname}%
\def\boldmath{\textfont1=\mibVIIIf\scriptfont1=\mibVIf
\scriptscriptfont1=\mibVf}%
\let\bfIXf=\bfVIIIf
 \if Y\REFEREE \normalbaselineskip=2\normalbaselineskip
 \normallineskip=2\normallineskip\fi
 \setbox\strutbox=\hbox{\vrule height7pt depth2pt width0pt}%
 \normalbaselines\rm}%
\def\begpet{\vskip\petitsurround
\bgroup\petit}
\def\endpet{\vskip\petitsurround
\egroup}

 \let  \tatss           = \bfXf
 \let  \tasss           = \syXf
 \let  \tbts            = \bfXf
 \let  \tbtss           = \bfVIIIf
 \let  \tbms            = \tamss
 \let  \tbmss           = \mibVIIIf
 \let  \tbss            = \syXf
 \let  \tbsss           = \syVIIIf
\def\newline{\hfill\break}
\def\rahmen#1{\vbox{\hrule\line{\vrule\vbox to#1true
cm{\vfil}\hfil\vrule}\vfil\hrule}}
\let\ts=\thinspace
\def\,{\relax\ifmmode\mskip\thinmuskip\else\thinspace\fi}
\def\unvskip{%
   \ifvmode
      \ifdim\lastskip=0pt
      \else
         \vskip-\lastskip
      \fi
   \fi}
\newtoks\eq\newtoks\eqn
\newdimen\mathhsize
\def\calcmathhsize{\mathhsize=\hsize
\advance\mathhsize by-\mathindent}
\calcmathhsize
\def\eqalign#1{\null\vcenter{\openup\jot\m@th
  \ialign{\strut\hfil$\displaystyle{##}$&$\displaystyle{{}##}$\hfil
      \crcr#1\crcr}}}
\def\displaylines#1{{}$\displ@y
\hbox{\vbox{\halign{$\@lign\hfil\displaystyle##\hfil$\crcr
    #1\crcr}}}${}}
\def\eqalignno#1{{}$\displ@y
  \hbox{\vbox{\halign
to\mathhsize{\hfil$\@lign\displaystyle{##}$\tabskip\z@skip
    &$\@lign\displaystyle{{}##}$\hfil\tabskip\centering
    &\llap{$\@lign##$}\tabskip\z@skip\crcr
    #1\crcr}}}${}}
\def\leqalignno#1{{}$\displ@y
\hbox{\vbox{\halign
to\mathhsize{\qquad\hfil$\@lign\displaystyle{##}$\tabskip\z@skip
    &$\@lign\displaystyle{{}##}$\hfil\tabskip\centering
    &\kern-\mathhsize\rlap{$\@lign##$}\tabskip\hsize\crcr
    #1\crcr}}}${}}
\def\generaldisplay{%
\ifeqno
       \ifleqno\leftline{$\displaystyle\the\eqn\quad\the\eq$}%
       \else\noindent\kern\mathindent\hbox to\mathhsize{$\displaystyle
             \the\eq\hfill\the\eqn$}%
       \fi
\else
       \kern\mathindent
       \hbox to\mathhsize{$\displaystyle\the\eq$\hss}%
\fi
\global\eq={}\global\eqn={}}%
\newif\ifeqno\newif\ifleqno
\everydisplay{\displaysetup}
\def\displaysetup#1$${\displaytest#1\eqno\eqno\displaytest}
\def\displaytest#1\eqno#2\eqno#3\displaytest{%
\if!#3!\ldisplaytest#1\leqno\leqno\ldisplaytest
\else\eqnotrue\leqnofalse\eqn={#2}\eq={#1}\fi
\generaldisplay$$}
\def\ldisplaytest#1\leqno#2\leqno#3\ldisplaytest{\eq={#1}%
\if!#3!\eqnofalse\else\eqnotrue\leqnotrue\eqn={#2}\fi}
\newcount\eqnum\eqnum=0
\def\autnum{\global\advance\eqnum by 1\relax{\rm(\the\eqnum)}}
\newdimen\lindent
\lindent=\stdparindent

\def\litemitem{\par\noindent\hbox to\lindent{\hfil}%
               \hangindent=2\lindent\ltextindent}
\def\ltextindent#1{\hbox to\lindent{#1\hss}\ignorespaces}
\def\set@item@mark#1{\llap{#1\enspace}\ignorespaces}
\ifx\undefined\mathhsize
   \def\item{\par\noindent
   \hangindent\itemindent\hangafter=0
   \set@item@mark}
   \def\itemitem{\par\noindent\advance\mathhsize by-\itemitemindent
   \hangindent\itemitemindent\hangafter=0
   \set@item@mark}
\else
   \def\item{\par\noindent\advance\mathhsize by-\itemindent
   \hangindent\itemindent\hangafter=0
   \everypar={\global\mathhsize=\hsize
   \global\advance\mathhsize by-\mathindent
   \global\everypar={}}\set@item@mark}
   \def\itemitem{\par\noindent\advance\mathhsize by-\itemitemindent
   \hangindent\itemitemindent\hangafter=0
   \everypar={\global\mathhsize=\hsize
   \global\advance\mathhsize by-\mathindent
   \global\everypar={}}\set@item@mark}
\fi
\newcount\the@end \global\the@end=0
\newbox\springer@macro \setbox\springer@macro=\vbox{}
\def\typeset{\setbox\springer@macro=\vbox{\begpet\noindent
   This article was processed by the author using
   Sprin\-ger-Ver\-lag \TeX{} A\&A macro package 1991.\par
   \egroup}\global\the@end=1}
\outer\def\bye{\bigskip\typeset
\sterne=1\ifx\speciali\undefined
\else
  \loop\smallskip\noindent special character No\number\sterne:
    \csname special\romannumeral\sterne\endcsname
    \advance\sterne by 1\relax
    \ifnum\sterne<11\relax
  \repeat
\fi
\if R\lr\null\fi\vfill\supereject\end}
\def\AALogo{\setbox254=\hbox{ ASTROPHYSICS }%
\vbox{\baselineskip=10dd\hrule\hbox{\vrule\vbox{\kern3pt
\hbox to\wd254{\hfil ASTRONOMY\hfil}
\hbox to\wd254{\hfil AND\hfil}\copy254
\hbox to\wd254{\hfil\number\day.\number\month.\number\year\hfil}
\kern3pt}\vrule}\hrule}}
\def\figure#1#2{\medskip\noindent{\petit{\bf Fig.\ts#1.\
}\ignorespaces#2\par}}
\def\tabcap#1#2{\smallskip\noindent{\bf Table\ts\ignorespaces
#1\unskip.\ }\ignorespaces #2\vskip3mm}
\expandafter \newcount \csname c@Tl\endcsname
    \csname c@Tl\endcsname=0
\expandafter \newcount \csname c@Tm\endcsname
    \csname c@Tm\endcsname=0
\expandafter \newcount \csname c@Tn\endcsname
    \csname c@Tn\endcsname=0
\expandafter \newcount \csname c@To\endcsname
    \csname c@To\endcsname=0
\expandafter \newcount \csname c@Tp\endcsname
    \csname c@Tp\endcsname=0
\expandafter \newcount \csname c@fn\endcsname
    \csname c@fn\endcsname=0
\def \stepc#1    {\global
    \expandafter
    \advance
    \csname c@#1\endcsname by 1}
\def \resetcount#1    {\global
    \csname c@#1\endcsname=0}
\def\@nameuse#1{\csname #1\endcsname}
\def\arabic#1{\@arabic{\@nameuse{c@#1}}}
\def\@arabic#1{\ifnum #1>0 \number #1\fi}
 \def \aTa  { \goodbreak
     \bgroup
     \par
 \textfont0=\tatt \scriptfont0=\tats \scriptscriptfont0=\tatss
 \textfont1=\tamt \scriptfont1=\tams \scriptscriptfont1=\tamss
 \textfont2=\tast \scriptfont2=\tass \scriptscriptfont2=\tasss
     \baselineskip=17dd\lineskiplimit=0pt\lineskip=0pt
     \rightskip=0pt plus4cm
     \pretolerance=10000
     \noindent
     \tatt}
 \def \eTa{\vskip10pt\egroup
     \noindent
     \ignorespaces}
 \def \aTb{\goodbreak
     \bgroup
     \par
 \textfont0=\tbtt \scriptfont0=\tbts \scriptscriptfont0=\tbtss
 \textfont1=\tbmt \scriptfont1=\tbms \scriptscriptfont1=\tbmss
 \textfont2=\tbst \scriptfont2=\tbss \scriptscriptfont2=\tbsss
     \baselineskip=13dd\lineskip=0pt\lineskiplimit=0pt
     \rightskip=0pt plus4cm
     \pretolerance=10000
     \noindent
     \tbtt}
 \def \eTb{\vskip10pt
     \egroup
     \noindent
     \ignorespaces}
\newcount\section@penalty  \section@penalty=0
\newcount\subsection@penalty  \subsection@penalty=0
\newcount\subsubsection@penalty  \subsubsection@penalty=0
\def\titlea#1{\par\stepc{Tl}
    \resetcount{Tm}
    \bgroup
       \normalsize
       \bf \rightskip 0pt plus4em
       \pretolerance=20000
       \boldmath
       \setbox0=\vbox{\vskip\tabefore
          \noindent
          \arabic{Tl}.\
          \ignorespaces#1
          \vskip\taafter}
       \dimen0=\ht0\advance\dimen0 by\dp0
       \advance\dimen0 by 2\baselineskip
       \advance\dimen0 by\pagetotal
       \ifdim\dimen0>\pagegoal
          \ifdim\pagetotal>\pagegoal
          \else\eject\fi\fi
       \vskip\tabefore
       \penalty\section@penalty \global\section@penalty=-200
       \global\subsection@penalty=10007
       \noindent
       \arabic{Tl}.\
       \ignorespaces#1
       \vskip\taafter
    \egroup
    \nobreak
    \parindent=0pt
    \let\lasttitle=A%
\everypar={\parindent=\stdparindent
    \penalty\z@\let\lasttitle=N\everypar={}}%
    \ignorespaces}
\def\titleb#1{\par\stepc{Tm}
    \resetcount{Tn}
    \if N\lasttitle\else\vskip\tbbeforeback\fi
    \bgroup
       \normalsize
       \raggedright
       \pretolerance=10000
       \it
       \setbox0=\vbox{\vskip\tbbefore
          \normalsize
          \raggedright
          \pretolerance=10000
          \noindent \it \arabic{Tl}.\arabic{Tm}.\ \ignorespaces#1
          \vskip\tbafter}
       \dimen0=\ht0\advance\dimen0 by\dp0\advance\dimen0 by 2\baselineskip
       \advance\dimen0 by\pagetotal
       \ifdim\dimen0>\pagegoal
          \ifdim\pagetotal>\pagegoal
          \else \if N\lasttitle\eject\fi \fi\fi
       \vskip\tbbefore
       \if N\lasttitle \penalty\subsection@penalty \fi
       \global\subsection@penalty=-100
       \global\subsubsection@penalty=10007
       \noindent \arabic{Tl}.\arabic{Tm}.\ \ignorespaces#1
       \vskip\tbafter
    \egroup
    \nobreak
    \let\lasttitle=B%
    \parindent=0pt
    \everypar={\parindent=\stdparindent
       \penalty\z@\let\lasttitle=N\everypar={}}%
       \ignorespaces}
\def\titlec#1{\par\stepc{Tn}
    \resetcount{To}
    \if N\lasttitle\else\vskip\tcbeforeback\fi
    \bgroup
       \normalsize
       \raggedright
       \pretolerance=10000
       \setbox0=\vbox{\vskip\tcbefore
          \noindent
          \arabic{Tl}.\arabic{Tm}.\arabic{Tn}.\
          \ignorespaces#1\vskip\tcafter}
       \dimen0=\ht0\advance\dimen0 by\dp0\advance\dimen0 by 2\baselineskip
       \advance\dimen0 by\pagetotal
       \ifdim\dimen0>\pagegoal
           \ifdim\pagetotal>\pagegoal
           \else \if N\lasttitle\eject\fi \fi\fi
       \vskip\tcbefore
       \if N\lasttitle \penalty\subsubsection@penalty \fi
       \global\subsubsection@penalty=-50
       \noindent
       \arabic{Tl}.\arabic{Tm}.\arabic{Tn}.\
       \ignorespaces#1\vskip\tcafter
    \egroup
    \nobreak
    \let\lasttitle=C%
    \parindent=0pt
    \everypar={\parindent=\stdparindent
       \penalty\z@\let\lasttitle=N\everypar={}}%
       \ignorespaces}
\def\titled#1{\par\stepc{To}
    \resetcount{Tp}
    \if N\lasttitle\else\vskip\tdbeforeback\fi
    \vskip\tdbefore
    \bgroup
       \normalsize
       \if N\lasttitle \penalty-50 \fi
       \it \noindent \ignorespaces#1\unskip\
    \egroup\ignorespaces}
\def\begref#1{\par
   \unvskip
   \goodbreak\vskip\tabefore
   {\noindent\bf\ignorespaces#1%
   \par\vskip\taafter}\nobreak\let\INS=N}
\def\ref{\if N\INS\let\INS=Y\else\goodbreak\fi
   \hangindent\stdparindent\hangafter=1\noindent\ignorespaces}
\def\endref{\goodbreak}
\def\acknow#1{\par
   \unvskip
   \vskip\tcbefore
   \noindent{\it Acknowledgements\/}. %
   \ignorespaces#1\par
   \vskip\tcafter}
\def\appendix#1{\vskip\tabefore
    \vbox{\noindent{\bf Appendix #1}\vskip\taafter}%
    \global\eqnum=0\relax
    \nobreak\noindent\ignorespaces}
\let\REFEREE=N
\newbox\refereebox
\setbox\refereebox=\vbox
to0pt{\vskip0.5cm\fullline{\hrulefill\tentt\lower0.5ex
\hbox{\kern5pt referee's copy\kern5pt}\hrulefill}\vss}%
\def\refereelayout{\let\REFEREE=M\footline={\copy\refereebox}
    \message{|A referee's copy will be produced}\par
    \if N\lr\else\if R\lr \onecolumn \fi \let\lr=N \topskip=10pt\fi}

\def\utw{\smash{\rlap{\lower5pt\hbox{$\sim$}}}}
\def\udtw{\smash{\rlap{\lower6pt\hbox{$\approx$}}}}


\def\bbbc{{\mathchoice {\setbox0=\hbox{$\displaystyle\rm C$}\hbox{\hbox
to0pt{\kern0.4\wd0\vrule height0.9\ht0\hss}\box0}}
{\setbox0=\hbox{$\textstyle\rm C$}\hbox{\hbox
to0pt{\kern0.4\wd0\vrule height0.9\ht0\hss}\box0}}
{\setbox0=\hbox{$\scriptstyle\rm C$}\hbox{\hbox
to0pt{\kern0.4\wd0\vrule height0.9\ht0\hss}\box0}}
{\setbox0=\hbox{$\scriptscriptstyle\rm C$}\hbox{\hbox
to0pt{\kern0.4\wd0\vrule height0.9\ht0\hss}\box0}}}}
\def\bbbq{{\mathchoice {\setbox0=\hbox{$\displaystyle\rm Q$}\hbox{\raise
0.15\ht0\hbox to0pt{\kern0.4\wd0\vrule height0.8\ht0\hss}\box0}}
{\setbox0=\hbox{$\textstyle\rm Q$}\hbox{\raise
0.15\ht0\hbox to0pt{\kern0.4\wd0\vrule height0.8\ht0\hss}\box0}}
{\setbox0=\hbox{$\scriptstyle\rm Q$}\hbox{\raise
0.15\ht0\hbox to0pt{\kern0.4\wd0\vrule height0.7\ht0\hss}\box0}}
{\setbox0=\hbox{$\scriptscriptstyle\rm Q$}\hbox{\raise
0.15\ht0\hbox to0pt{\kern0.4\wd0\vrule height0.7\ht0\hss}\box0}}}}
\def\bbbt{{\mathchoice {\setbox0=\hbox{$\displaystyle\rm
T$}\hbox{\hbox to0pt{\kern0.3\wd0\vrule height0.9\ht0\hss}\box0}}
{\setbox0=\hbox{$\textstyle\rm T$}\hbox{\hbox
to0pt{\kern0.3\wd0\vrule height0.9\ht0\hss}\box0}}
{\setbox0=\hbox{$\scriptstyle\rm T$}\hbox{\hbox
to0pt{\kern0.3\wd0\vrule height0.9\ht0\hss}\box0}}
{\setbox0=\hbox{$\scriptscriptstyle\rm T$}\hbox{\hbox
to0pt{\kern0.3\wd0\vrule height0.9\ht0\hss}\box0}}}}
\def\bbbs{{\mathchoice
{\setbox0=\hbox{$\displaystyle     \rm S$}\hbox{\raise0.5\ht0\hbox
to0pt{\kern0.35\wd0\vrule height0.45\ht0\hss}\hbox
to0pt{\kern0.55\wd0\vrule height0.5\ht0\hss}\box0}}
{\setbox0=\hbox{$\textstyle        \rm S$}\hbox{\raise0.5\ht0\hbox
to0pt{\kern0.35\wd0\vrule height0.45\ht0\hss}\hbox
to0pt{\kern0.55\wd0\vrule height0.5\ht0\hss}\box0}}
{\setbox0=\hbox{$\scriptstyle      \rm S$}\hbox{\raise0.5\ht0\hbox
to0pt{\kern0.35\wd0\vrule height0.45\ht0\hss}\raise0.05\ht0\hbox
to0pt{\kern0.5\wd0\vrule height0.45\ht0\hss}\box0}}
{\setbox0=\hbox{$\scriptscriptstyle\rm S$}\hbox{\raise0.5\ht0\hbox
to0pt{\kern0.4\wd0\vrule height0.45\ht0\hss}\raise0.05\ht0\hbox
to0pt{\kern0.55\wd0\vrule height0.45\ht0\hss}\box0}}}}
\def\bbbz{{\mathchoice {\hbox{$\sf\textstyle Z\kern-0.4em Z$}}
{\hbox{$\sf\textstyle Z\kern-0.4em Z$}}
{\hbox{$\sf\scriptstyle Z\kern-0.3em Z$}}
{\hbox{$\sf\scriptscriptstyle Z\kern-0.2em Z$}}}}
\def\diameter{{\ifmmode\oslash\else$\oslash$\fi}}

\def\vec#1{{\boldmath
\textfont0=\bfIXf\scriptfont0=\bfVIf\scriptscriptfont0=\bfVf
\ifmmode
\mathchoice{\hbox{$\displaystyle#1$}}{\hbox{$\textstyle#1$}}
{\hbox{$\scriptstyle#1$}}{\hbox{$\scriptscriptstyle#1$}}\else
$#1$\fi}}
\def\tens#1{\ifmmode
\mathchoice{\hbox{$\displaystyle\sf#1$}}{\hbox{$\textstyle\sf#1$}}
{\hbox{$\scriptstyle\sf#1$}}{\hbox{$\scriptscriptstyle\sf#1$}}\else
$\sf#1$\fi}
\newcount\sterne \sterne=0
\newdimen\fullhead
{\catcode`@=11    
\def\newtoks{\alloc@5\toks\toksdef\@cclvi}
\outer\gdef\makenewtoks#1{\newtoks#1#1={ ????? }}}
\makenewtoks\DATE
\makenewtoks\MAINTITLE
\makenewtoks\SUBTITLE
\makenewtoks\AUTHOR
\makenewtoks\INSTITUTE
\makenewtoks\ABSTRACT
\makenewtoks\KEYWORDS
\makenewtoks\THESAURUS
\makenewtoks\OFFPRINTS
\newlinechar=`\| %
\let\INS=N%
{\catcode`\@=\active
\gdef@#1{\if N\INS $^{#1}$\else\if
E\INS\hangindent0.5\stdparindent\hangafter=1%
\noindent\hbox to0.5\stdparindent{$^{#1}$\hfil}\let\INS=Y\ignorespaces
\else\par\hangindent0.5\stdparindent\hangafter=1
\noindent\hbox to0.5\stdparindent{$^{#1}$\hfil}\ignorespaces\fi\fi}%
}%
\def\mehrsterne{\global\advance\sterne by1\relax}%
\def\footnoterule{\kern-3pt\hrule width 2true cm\kern2.6pt}
\def\makeOFFPRINTS#1{\bgroup\normalsize
       \hsize=19.5cc
       \baselineskip=10dd\lineskiplimit=0pt\lineskip=0pt
       \def\textindent##1{\noindent{\it Send offprint
          requests to\/}: }\relax
       \vfootnote{nix}{\ignorespaces#1}\egroup}
\def\makesterne{\count254=0\loop\ifnum\count254<\sterne
\advance\count254 by1\star\repeat}
\def\FOOTNOTE#1{\bgroup
       \ifhmode\unskip\fi
       \mehrsterne$^{\makesterne}$\relax
       \normalsize
       \hsize=19.5cc
       \baselineskip=10dd\lineskiplimit=0pt\lineskip=0pt
       \def\textindent##1{\noindent\hbox
       to\stdparindent{##1\hss}}\relax
       \vfootnote{$^{\makesterne}$}{\ignorespaces#1}\egroup}
\def\fonote#1{\ifhmode\unskip\fi
       \mehrsterne$^{\the\sterne}$\bgroup
       \normalsize
       \hsize=19.5cc
       \def\textindent##1{\noindent\hbox
       to\stdparindent{##1\hss}}\relax
       \vfootnote{$^{\the\sterne}$}{\ignorespaces#1}\egroup}
\def\missmsg#1{\message{|Missing #1 }}
\def\tstmiss#1#2#3#4#5{%
\edef\test{\the #1}%
\ifx\test\missing%
  #2\relax
  #3
\else
  \ifx\test\missingi%
    #2\relax
    #3
  \else #4
  \fi
\fi
#5
}%
\def\maketitle{\paglay%
\def\missing{ ????? }%
\def\missingi{ }%
{\parskip=0pt\relax
\setbox0=\vbox{\hsize=\fullhsize\null\vskip2truecm
\tstmiss%
  {\MAINTITLE}%
  {}%
  {\global\MAINTITLE={MAINTITLE should be given}}%
  {}%
  {
   \aTa\ignorespaces\the\MAINTITLE\eTa}%
\tstmiss%
  {\SUBTITLE}%
  {}%
  {}%
  {
   \aTb\ignorespaces\the\SUBTITLE\eTb}%
  {}%
\tstmiss%
  {\AUTHOR}%
  {}%
  {\AUTHOR={Name(s) and initial(s) of author(s) should be given}}
  {}%
  {
\noindent{\bf\ignorespaces\the\AUTHOR\vskip4pt}}%
\tstmiss%
  {\INSTITUTE}%
  {}%
  {\INSTITUTE={Address(es) of author(s) should be given.}}%
  {}%
  {
   \let\INS=E
\noindent\ignorespaces\the\INSTITUTE\vskip10pt}%
\tstmiss%
  {\DATE}%
  {}%
  {\DATE={$[$the date of receipt and acceptance should be inserted
later$]$}}%
  {}%
  {
{\noindent\ignorespaces\the\DATE\vskip21pt}\bf A}%
}%
\global\fullhead=\ht0\global\advance\fullhead by\dp0
\global\advance\fullhead by10pt\global\sterne=0
{\hsize=19.5cc\null\vskip2truecm
\tstmiss%
  {\OFFPRINTS}%
  {}%
  {}%
  {\makeOFFPRINTS{\the\OFFPRINTS}}%
  {}%
\hsize=\fullhsize
\tstmiss%
  {\MAINTITLE}%
  {\missmsg{MAINTITLE}}%
  {\global\MAINTITLE={MAINTITLE should be given}}%
  {}%
  {
   \aTa\ignorespaces\the\MAINTITLE\eTa}%
\tstmiss%
  {\SUBTITLE}%
  {}%
  {}%
  {
   \aTb\ignorespaces\the\SUBTITLE\eTb}%
  {}%
\tstmiss%
  {\AUTHOR}%
  {\missmsg{name(s) and initial(s) of author(s)}}%
  {\AUTHOR={Name(s) and initial(s) of author(s) should be given}}
  {}%
  {
\noindent{\bf\ignorespaces\the\AUTHOR\vskip4pt}}%
\tstmiss%
  {\INSTITUTE}%
  {\missmsg{address(es) of author(s)}}%
  {\INSTITUTE={Address(es) of author(s) should be given.}}%
  {}%
  {
   \let\INS=E
\noindent\ignorespaces\the\INSTITUTE\vskip10pt}%
\catcode`\@=12
\tstmiss%
  {\DATE}%
  {\message{|The date of receipt and acceptance should be inserted
later.}}%
  {\DATE={$[$the date of receipt and acceptance should be inserted
later$]$}}%
  {}%
  {
{\noindent\ignorespaces\the\DATE\vskip21pt}}%
}%
\tstmiss%
  {\THESAURUS}%
  {\message{|Thesaurus codes are not given.}}%
  {\global\THESAURUS={missing; you have not inserted them}}%
  {}%
  {}%
\if M\REFEREE\let\REFEREE=Y
\normalbaselineskip=2\normalbaselineskip
\normallineskip=2\normallineskip\normalbaselines\fi
\tstmiss%
  {\ABSTRACT}%
  {\missmsg{ABSTRACT}}%
  {\ABSTRACT={Not yet given.}}%
  {}%
  {\noindent{\bf Abstract. }\ignorespaces\the\ABSTRACT\vskip0.5true cm}%
\def\strich{\par
\vbox to0pt{\hrule width\hsize\vss}\vskip-1.2\baselineskip
\vskip0pt plus3\baselineskip\relax}%
\tstmiss%
  {\KEYWORDS}%
  {\missmsg{KEYWORDS}}%
  {\KEYWORDS={Not yet given.}}%
  {}%
  {\noindent{\bf Key words: }\ignorespaces\the\KEYWORDS
  \strich}%
\global\sterne=0
}}
\newdimen\@txtwd  \@txtwd=\hsize
\newdimen\@txtht  \@txtht=\vsize
\newdimen\@colht  \@colht=\vsize
\newdimen\@colwd  \@colwd=-1pt
\newdimen\@colsavwd
\newcount\in@t \in@t=0
\def\initlr{\if N\lr \ifdim\@colwd<0pt \global\@colwd=\hsize \fi
   \else\global\let\lr=L\ifdim\@colwd<0pt \global\@colwd=\hsize
      \global\divide\@colwd\tw@ \global\advance\@colwd by -10pt
   \fi\fi\global\advance\in@t by 1}
\def\setuplr#1#2#3{\let\lr=O \ifx#1\lr\global\let\lr=N
      \else\global\let\lr=L\fi
   \@txtht=\vsize \@colht=\vsize \@txtwd=#2 \@colwd=#3
   \if N\lr \else\multiply\@colwd\tw@ \fi
   \ifdim\@colwd>\@txtwd\if N\lr
        \errmessage{The text width is less than the column width}%
      \else
        \errmessage{The text width is less the two times the column width}%
      \fi \global\@colwd=\@txtwd
      \if N\lr\divide\@colwd by 2\fi
   \else \global\@colwd=#3 \fi \initlr \@colsavwd=#3
   \global\@insmx=\@txtht
   \global\hsize=\@colwd}
\def\twocolumns{\@fillpage\eject\global\let\lr=L \@makecolht
   \global\@colwd=\@colsavwd \global\hsize=\@colwd}
\def\onecolumn{\@fillpage\eject\global\let\lr=N \@makecolht
   \global\@colwd=\@txtwd \global\hsize=\@colwd}
\def\newpage{\@fillpage\eject}
\def\@fillpage{\vfill\supereject\if R\lr \null\vfill\eject\fi}

\newbox\@leftcolumn
\newbox\@rightcolumn
\newbox\@outputbox
\newbox\@tempboxa
\newbox\@keepboxa
\newbox\@keepboxb
\newbox\@bothcolumns
\newbox\@savetopins
\newbox\@savetopright
\newcount\verybad \verybad=1010
\def\@makecolumn{\ifnum \in@t<1\initlr\fi
   \ifnum\outputpenalty=\the\verybad1  
      \if L\lr\else\advance\pageno by1\fi
      \message{Warning: There is a 'widow' line
      at the top of page \the\pageno\if R\lr (left)\fi.
      This is unacceptable.} \if L\lr\else\advance\pageno by-1\fi \fi
   \ifnum\outputpenalty=\the\verybad2
      \message{Warning: There is a 'club' line
      at the bottom of page \the\pageno\if L\lr(left)\fi.
      This is unacceptable.} \fi
   \if L\lr \ifvoid\@savetopins\else\@colht=\@txtht\fi \fi
   \if R\lr \ifvoid\@bothcolumns \ifvoid\@savetopright
       \else\@colht=\@txtht\fi\fi\fi
   \global\setbox\@outputbox
   \vbox to\@colht{\boxmaxdepth\maxdepth
   \if L\lr \ifvoid\@savetopins\else\unvbox\@savetopins\fi \fi
   \if R\lr \ifvoid\@bothcolumns \ifvoid\@savetopright\else
       \unvbox\@savetopright\fi\fi\fi
   \ifvoid\topins\else\ifnum\count\topins>0
         \ifdim\ht\topins>\@colht
            \message{|Error: Too many or too large single column
            box(es) on this page.}\fi
         \unvbox\topins
      \else
         \global\setbox\@savetopins=\vbox{\ifvoid\@savetopins\else
         \unvbox\@savetopins\penalty-500\fi \unvbox\topins} \fi\fi
   \dimen@=\dp\@cclv \unvbox\@cclv 
   \ifvoid\bottomins\else\unvbox\bottomins\fi
   \ifvoid\footins\else 
     \vskip\skip\footins
     \footnoterule
     \unvbox\footins\fi
   \ifr@ggedbottom \kern-\dimen@ \vfil \fi}%
}
\def\@outputpage{\@dooutput{\lr}}
\def\@colbox#1{\hbox to\@colwd{\box#1\hss}}
\def\@dooutput#1{\global\topskip=10pt
  \ifdim\ht\@bothcolumns>\@txtht
    \if #1N
       \unvbox\@outputbox
    \else
       \unvbox\@leftcolumn\unvbox\@outputbox
    \fi
    \global\setbox\@tempboxa\vbox{\hsize=\@txtwd\makeheadline
       \vsplit\@bothcolumns to\@txtht
       \makefootline\hsize=\@colwd}%
    \message{|Error: Too many double column boxes on this page.}%
    \shipout\box\@tempboxa\advancepageno
    \unvbox255 \penalty\outputpenalty
  \else
    \global\setbox\@tempboxa\vbox{\hsize=\@txtwd\makeheadline
       \ifvoid\@bothcolumns\else\unvbox\@bothcolumns\fi
       \hsize=\@colwd
       \if #1N
          \hbox to\@txtwd{\@colbox{\@outputbox}\hfil}%
       \else
          \hbox to\@txtwd{\@colbox{\@leftcolumn}\hfil\@colbox{\@outputbox}}%
       \fi
       \hsize=\@txtwd\makefootline\hsize=\@colwd}%
    \shipout\box\@tempboxa\advancepageno
  \fi
  \ifnum \special@pages>0 \s@count=100 \page@command
      \xdef\page@command{}\global\special@pages=0 \fi
  }
\def\balance@right@left{\dimen@=\ht\@leftcolumn
    \advance\dimen@ by\ht\@outputbox
    \advance\dimen@ by\ht\springer@macro
    \dimen2=\z@ \global\the@end=0
    \ifdim\dimen@>70pt\setbox\z@=\vbox{\unvbox\@leftcolumn
          \unvbox\@outputbox}%
       \loop
          \dimen@=\ht\z@
          \advance\dimen@ by0.5\topskip
          \advance\dimen@ by\baselineskip
          \advance\dimen@ by\ht\springer@macro
          \advance\dimen@ by\dimen2
          \divide\dimen@ by2
          \splittopskip=\topskip
          {\vbadness=10000
             \global\setbox3=\copy\z@
             \global\setbox1=\vsplit3 to \dimen@}%
          \dimen1=\ht3 \advance\dimen1 by\ht\springer@macro
       \ifdim\dimen1>\ht1 \advance\dimen2 by\baselineskip\repeat
       \dimen@=\ht1
       \global\setbox\@leftcolumn
          \hbox to\@colwd{\vbox to\@colht{\vbox to\dimen@{\unvbox1}\vfil}}%
       \global\setbox\@outputbox
          \hbox to\@colwd{\vbox to\@colht{\vbox to\dimen@{\unvbox3
             \vfill\box\springer@macro}\vfil}}%
    \else
       \setbox\@leftcolumn=\vbox{unvbox\@leftcolumn\bigskip
          \box\springer@macro}%
    \fi}
\newinsert\bothins
\newbox\rightins
\skip\bothins=\z@skip
\count\bothins=1000
\dimen\bothins=\@txtht \advance\dimen\bothins by -\bigskipamount
\def\bothtopinsert{\par\begingroup\setbox\z@\vbox\bgroup
    \hsize=\@txtwd\parskip=0pt\par\noindent\bgroup}
\def\endbothinsert{\egroup\egroup
  \if R\lr
    \right@nsert
  \else    
    \dimen@=\ht\z@ \advance\dimen@ by\dp\z@ \advance\dimen@ by\pagetotal
    \advance\dimen@ by \bigskipamount \advance\dimen@ by \topskip
    \advance\dimen@ by\ht\topins \advance\dimen@ by\dp\topins
    \advance\dimen@ by\ht\bottomins \advance\dimen@ by\dp\bottomins
    \advance\dimen@ by\ht\@savetopins \advance\dimen@ by\dp\@savetopins
    \ifdim\dimen@>\@colht\right@nsert\else\left@nsert\fi
  \fi  \endgroup}
\def\right@nsert{\global\setbox\rightins\vbox{\ifvoid\rightins
    \else\unvbox\rightins\fi\penalty100
    \splittopskip=\topskip
    \splitmaxdepth\maxdimen \floatingpenalty200
    \dimen@\ht\z@ \advance\dimen@\dp\z@
    \box\z@\nobreak\bigskip}}
\def\left@nsert{\insert\bothins{\penalty100
    \splittopskip=\topskip
    \splitmaxdepth\maxdimen \floatingpenalty200
    \box\z@\nobreak\bigskip}
    \@makecolht}
\newdimen\@insht    \@insht=\z@
\newdimen\@insmx    \@insmx=\vsize
\def\@makecolht{\global\@colht=\@txtht \@compinsht
    \global\advance\@colht by -\@insht \global\vsize=\@colht
    \global\dimen\topins=\@colht}
\def\@compinsht{\if R\lr
       \dimen@=\ht\@bothcolumns \advance\dimen@ by\dp\@bothcolumns
       \ifvoid\@bothcolumns \advance\dimen@ by\ht\@savetopright
          \advance\dimen@ by\dp\@savetopright \fi
    \else
       \dimen@=\ht\bothins \advance\dimen@ by\dp\bothins
       \advance\dimen@ by\ht\@savetopins \advance\dimen@ by\dp\@savetopins
    \fi
    \ifdim\dimen@>\@insmx
       \global\@insht=\dimen@
    \else\global\@insht=\dimen@
    \fi}
\newinsert\bottomins
\skip\bottomins=\z@skip
\count\bottomins=1000
\xdef\page@command{}
\newcount\s@count
\newcount\special@pages \special@pages=0
\def\specialpage#1{\global\advance\special@pages by1
    \global\s@count=\special@pages
    \global\advance\s@count by 100
    \global\setbox\s@count
    \vbox to\@txtht{\hsize=\@txtwd\parskip=0pt
    \par\noindent\noexpand#1\vfil}%
    \def\protect{\noexpand\protect\noexpand}%
    \xdef\page@command{\page@command
         \protect\global\advance\s@count by1
         \protect\begingroup
         \protect\setbox\z@\vbox{\protect\makeheadline
                                    \protect\box\s@count
            \protect\makefootline}%
         \protect{\shipout\box\z@}%
         \protect\endgroup\protect\advancepageno}%
    \let\protect=\relax
   }
\def\@startins{\vskip \topskip\hrule height\z@
   \nobreak\vskip -\topskip\vskip3.7pt}
\let\retry=N
\output={\@makecolht \global\topskip=10pt \let\retry=N%
   \ifnum\count\topins>0 \ifdim\ht\topins>\@colht
       \global\count\topins=0 \global\let\retry=Y%
       \unvbox\@cclv \penalty\outputpenalty \fi\fi
   \if N\retry
    \if N\lr     
       \@makecolumn
       \ifnum\the@end>0
          \setbox\z@=\vbox{\unvcopy\@outputbox}%
          \dimen@=\ht\z@ \advance\dimen@ by\ht\springer@macro
          \ifdim\dimen@<\@colht
             \setbox\@outputbox=\vbox to\@colht{\box\z@
             \unskip\vskip12pt plus0pt minus12pt
             \box\springer@macro\vfil}%
          \else \box\springer@macro \fi
          \global\the@end=0
       \fi
       \ifvoid\bothins\else\global\setbox\@bothcolumns\box\bothins\fi
       \@outputpage
       \ifvoid\rightins\else
       \ifvoid\@savetopins\insert\bothins{\unvbox\rightins}\fi
       \fi
    \else
       \if L\lr    
          \@makecolumn
          \global\setbox\@leftcolumn\box\@outputbox \global\let\lr=R%
          \ifnum\pageno=1
             \message{|[left\the\pageno]}%
          \else
             \message{[left\the\pageno]}\fi
          \ifvoid\bothins\else\global\setbox\@bothcolumns\box\bothins\fi
          \global\dimen\bothins=\z@
          \global\count\bothins=0
          \ifnum\pageno=1
             \global\topskip=\fullhead\fi
       \else    
          \@makecolumn
          \ifnum\the@end>0\ifnum\pageno>1\balance@right@left\fi\fi
          \@outputpage \global\let\lr=L%
          \global\dimen\bothins=\maxdimen
          \global\count\bothins=1000
          \ifvoid\rightins\else
             \ifvoid\@savetopins \insert\bothins{\unvbox\rightins}\fi
          \fi
       \fi
    \fi
    \global\let\last@insert=N \put@default
    \ifnum\outputpenalty>-\@MM\else\dosupereject\fi
    \ifvoid\@savetopins\else
      \ifdim\ht\@savetopins>\@txtht
        \global\setbox\@tempboxa=\box\@savetopins
        \global\setbox\@savetopins=\vsplit\@tempboxa to\@txtht
        \global\setbox\@savetopins=\vbox{\unvbox\@savetopins}%
        \global\setbox\@savetopright=\box\@tempboxa \fi
    \fi
    \@makecolht
    \global\count\topins=1000
   \fi
   }
\if N\lr
   \setuplr{O}{\fullhsize}{\hsize}
\else
   \setuplr{T}{\fullhsize}{\hsize}
\fi
\def\put@default{\global\let\insert@here=Y
   \global\let\insert@at@the@bottom=N}%
\def\puthere{\global\let\insert@here=Y%
    \global\let\insert@at@the@bottom=N}
\def\putattop{\global\let\insert@here=N%
    \global\let\insert@at@the@bottom=N}
\def\putatbottom{\global\let\insert@here=N%
    \global\let\insert@at@the@bottom=X}
\put@default
\let\last@insert=N
\def\end@skip{\smallskip}
\newdimen\min@top
\newdimen\min@here
\newdimen\min@bot
\min@top=10cm
\min@here=4cm
\min@bot=\topskip
\def\figfuzz{\vskip 0pt plus 6pt minus 3pt}  
\def\check@here@and@bottom#1{\relax
   \ifvoid\topins\else       \global\let\insert@here=N\fi
   \if B\last@insert         \global\let\insert@here=N\fi
   \if T\last@insert         \global\let\insert@here=N\fi
   \ifdim #1<\min@bot        \global\let\insert@here=N\fi
   \ifdim\pagetotal>\@colht  \global\let\insert@here=N\fi
   \ifdim\pagetotal<\min@here\global\let\insert@here=N\fi
   \if X\insert@at@the@bottom\global\let\insert@at@the@bottom=Y
     \else\if T\last@insert  \global\let\insert@at@the@bottom=N\fi
          \if H\last@insert  \global\let\insert@at@the@bottom=N\fi
          \ifvoid\topins\else\global\let\insert@at@the@bottom=N\fi\fi
   \ifdim #1<\min@bot        \global\let\insert@at@the@bottom=N\fi
   \ifdim\pagetotal>\@colht  \global\let\insert@at@the@bottom=N\fi
   \ifdim\pagetotal<\min@top \global\let\insert@at@the@bottom=N\fi
   \ifvoid\bottomins\else    \global\let\insert@at@the@bottom=Y\fi
   \if Y\insert@at@the@bottom\global\let\insert@here=N\fi }
\def\single@column@insert#1{\relax
   \setbox\@tempboxa=\vbox{#1}%
   \dimen@=\@colht \advance\dimen@ by -\pagetotal
   \advance\dimen@ by-\ht\@tempboxa \advance\dimen0 by-\dp\@tempboxa
   \advance\dimen@ by-\ht\topins \advance\dimen0 by-\dp\topins
   \check@here@and@bottom{\dimen@}%
   \if Y\insert@here
      \par  
      \midinsert\figfuzz\relax     
      \box\@tempboxa\end@skip\figfuzz\endinsert
      \global\let\last@insert=H
   \else \if Y\insert@at@the@bottom
      \begingroup\insert\bottomins\bgroup\if B\last@insert\end@skip\fi
      \floatingpenalty=20000\figfuzz\bigskip\box\@tempboxa\egroup\endgroup
      \global\let\last@insert=B
   \else
      \topinsert\box\@tempboxa\end@skip\figfuzz\endinsert
      \global\let\last@insert=T
   \fi\fi\put@default\ignorespaces}
\def\begfig#1cm#2\endfig{\single@column@insert{\@startins\rahmen{#1}#2}%
\ignorespaces}
\def\begfigwid#1cm#2\endfig{\relax
   \if N\lr  
      {\hsize=\fullhsize \begfig#1cm#2\endfig}%
   \else
      \setbox0=\vbox{\hsize=\fullhsize\bigskip#2\smallskip}%
      \dimen0=\ht0\advance\dimen0 by\dp0
      \advance\dimen0 by#1cm
      \advance\dimen0by7\normalbaselineskip\relax
      \ifdim\dimen0>\@txtht
         \message{|Figure plus legend too high, will try to put it on a
                  separate page. }%
         \begfigpage#1cm#2\endfig
      \else
         \bothtopinsert\line{\vbox{\hsize=\fullhsize
         \@startins\rahmen{#1}#2\smallskip}\hss}\figfuzz\endbothinsert
      \fi
   \fi}
\def\begfigside#1cm#2cm#3\endfig{\relax
   \if N\lr  
      {\hsize=\fullhsize \begfig#1cm#3\endfig}%
   \else
      \dimen0=#2true cm\relax
      \ifdim\dimen0<\hsize
         \message{|Your figure fits in a single column; why don't|you use
                  \string\begfig\space instead of \string\begfigside? }%
      \fi
      \dimen0=\fullhsize
      \advance\dimen0 by-#2true cm
      \advance\dimen0 by-1true cc\relax
      \bgroup
         \ifdim\dimen0<8true cc\relax
            \message{|No sufficient room for the legend;
                     using \string\begfigwid. }%
            \begfigwid #1cm#3\endfig
         \else
            \ifdim\dimen0<10true cc\relax
               \message{|Room for legend to narrow;
                        legend will be set raggedright. }%
               \rightskip=0pt plus 2cm\relax
            \fi
            \setbox0=\vbox{\def\figure##1##2{\vbox{\hsize=\dimen0\relax
                           \@startins\noindent\petit{\bf
                           Fig.\ts##1\unskip.\ }\ignorespaces##2\par}}%
                           #3\unskip}%
            \ifdim#1true cm<\ht0\relax
               \message{|Text of legend higher than figure; using
                        \string\begfig. }%
               \begfigwid #1cm#3\endfig
            \else
               \def\figure##1##2{\vbox{\hsize=\dimen0\relax
                                       \@startins\noindent\petit{\bf
                                       Fig.\ts##1\unskip.\
                                       }\ignorespaces##2\par}}%
               \bothtopinsert\line{\vbox{\hsize=#2true cm\relax
               \@startins\rahmen{#1}}\hss#3\unskip}\figfuzz\endbothinsert
            \fi
         \fi
      \egroup
   \fi\ignorespaces}
\def\begfigpage#1cm#2\endfig{\specialpage{\@startins
   \vskip3.7pt\rahmen{#1}#2}\ignorespaces}%
\def\begtab#1cm#2\endtab{\single@column@insert{#2\rahmen{#1}}\ignorespaces}
\let\begtabempty=\begtab
\def\begtabfull#1\endtab{\single@column@insert{#1}\ignorespaces}
\def\begtabemptywid#1cm#2\endtab{\relax
   \if N\lr
      {\hsize=\fullhsize \begtabempty#1cm#2\endtab}%
   \else
      \bothtopinsert\line{\vbox{\hsize=\fullhsize
      #2\rahmen{#1}}\hss}\medskip\endbothinsert
   \fi\ignorespaces}
\def\begtabfullwid#1\endtab{\relax
   \if N\lr
      {\hsize=\fullhsize \begtabfull#1\endtab}%
   \else
      \bothtopinsert\line{\vbox{\hsize=\fullhsize
      \noindent#1}\hss}\medskip\endbothinsert
   \fi\ignorespaces}
\def\begtabpage#1\endtab{\specialpage{#1}\ignorespaces}
\catcode`\@=\active   

\def\cbrt#1{\root 3 \of {#1}}
\def\tree{{\SCS Tree}}
%
%
  \MAINTITLE={On the evolution of shape in N--body simulations}
  \SUBTITLE={}
  \AUTHOR={ Ch.\ Theis$^1$, R. Spurzem$^2$}
  \OFFPRINTS={Ch.\ Theis}
 \INSTITUTE={${}^1$ Inst. f"ur Theoretische Physik und Astrophysik, Univ. Kiel,
                        Olshausenstra\3e 40, D-24098 Kiel, Germany

             \noindent
             ${}^2$ Astronomisches Rechen--Institut,  
                 M"onchhofstr.\ 12--14, 69120 Heidelberg, Germany}
%
%
 \DATE={ Received; accepted }
%
%
 \ABSTRACT={A database on shape evolution of direct $N$-body models
 formed out of cold, dissipationless collapse is generated using GRAPE
 and HARP special purpose computers. Such models are important
 to understand the formation of elliptical galaxies.
 Three dynamically distinct phases
 of shape evolution were found, first a fast dynamical collapse which
 gives rise to the radial orbit instability (ROI) and generates at its
 end the maximal
 triaxiality of the system. Subsequently, two phases of violent and two-body
 shape relaxation occur, which drive the system first towards axisymmetry,
 finally to spherical symmetry (the final state, however, is still much
 more concentrated than the initial model). 
 In a sequence of models the influence of numerical
 and physical parameters, like particle number, softening, initial virial
 ratio, timestep choice, different $N$-body codes, are examined. We find
 that an improper combination of softening and particle number can produce
 erroneous results. Selected models were evolved on the secular timescale
 until they became spherically symmetric again. The secular shape relaxation
 time scale is shown to agree very well with the two-body relaxation time,
 if softening is properly taken into account for the latter. Finally, we
 argue, that the intermediate phase of violent shape relaxation after
 collapse is induced by strong core oscillations in the centre, which cause
 potential fluctuations, dampening out the triaxiality.
   }
%
%
%
   \KEYWORDS={Numerical methods -- Galaxies: kinematics and dynamics
    -- Galaxies: evolution
    }
  \THESAURUS={14.03.1,07.14.1,07.07.1}
%
\maketitle
 \titlea{Introduction}
 Formation scenarios of elliptical galaxies are either based on merging of
 galaxies or on violently collapsing stellar systems (or on 
 combinations of both). Especially the field ellipticals
 most likely pass through a phase of cold dissipationless collapse. 
 It only occurs if the primordial gas cloud forming a galaxy
 has a very efficient phase of star formation at the beginning
 by which practically all the gas is transformed into stars.
 Thereafter
 the system is not supported against further collapse until a 
 dynamical equilibrium is reached. The
 structure of the final state is restricted by the time-independent
 tensor-virial theorem (see Binney \& Tremaine 1987), which
 includes the possibility of anisotropy-supported triaxial configurations. 
 
 Merritt \& Aguilar (1985) and Min \& Choi (1989)
 have shown by $N$-body simulations that the initially spherical
 system does not remain spherical during collapse.
 If approximately $2T/|U| < 0.1$, where $T$ and $U$ are
 the kinetic and potential energy of the initial state, the
 system is unstable against the formation of a nearly prolate shape.
 Aguilar \& Merritt (1990) performed an exhaustive study of
 the final shape of the prolate configuration as a function
 of initial parameters as kinetic energy and initial triaxiality.
 Cannizzo \& Hollister (1992) varied in a similar parameter study
 the initial density profile and the softening parameter (three
 values).

 This instability, commonly known as ``radial orbit instability''
 (henceforth ROI) was long believed to be connected with some initial
 amount of anisotropy in the velocity dispersions of the system,
 which either stems from the initial conditions or develops during
 collapse. However, Udry (1993) showed that even a very small
 initial anisotropy suffices to develop ROI,
 and more recently Hozumi, Fujiwara
 \& Kan-ya (1996) pointed out that the onset of ROI happens
 during the collapse of a uniform sphere while it remains completely
 isotropic. So, although it is usually stated that the onset of ROI
 can be understood as a type of Jeans instability
 perpendicular to the radial movement of collapse,
 a complete understanding of its mechanism is still lacking.
 It is interesting to note, however, that some remarks made by
 Bettwieser \& Spurzem (1986) and Hensler et al. (1995) can be
 applied here. They showed that homogeneous systems in self-similar
 collapse remain exactly isotropic, whereas in other cases anisotropy
 develops proportional to the gradient of $u/r$, where $u$ is the
 radial bulk mass streaming velocity.

 ROI is related to the concept of global instability of stellar systems, 
 although details of this relation are still unclear.
 The global instability of a spherical system of
 purely radially moving stars had been proven by
 Antonov (1961) and Polyachenko \& Shukhman (1981) analytically.
 Most other analytic or semi-analytic stability criteria rely
 on an energy principle, but can be applied only to systems where
 the distribution function is a function of energy alone, i.e.
 isotropic systems (Antonov 1961, 1962, Lynden-Bell 1964,
 Ipser \& Thorne 1968, Kulsrud \& Mark 1970).
 For most anisotropic and
 more realistic cases, however, no rigorous proof of sufficient
 or necessary criteria for stability
 is known, and one has to rely on numerical simulations, as in the
 papers cited in the previous paragraphs.

 Lynden-Bell (1967) derived by using principles of statistical mechanics
 a most probable distribution function as the end-state of cold, 
 dissipationless collapse, the
 Lynden-Bell distribution. The assumed nature of the
 process led to the notion of violent relaxation.
 However, already very early
 $N$-body simulations (Cuperman, Goldstein \& Lecar 1969) were not
 able to show that such a distribution occurs. Burkert (1990)
 tried to relax the problem by assuming that the relaxation
 is incomplete, i.e. not all elements of phase space are accessible
 with equal probability, but the concept of violent relaxation
 thereby became not so attractive anymore. There was a remarkable
 series of papers (Wiechen, Ziegler \& Schindler 1988,
 Ziegler \& Wiechen 1989, 1990),
 who derived the final state as a function of the initial state by
 an energy principle. However, it has not yet been practically
 applied, as far as the authors of this paper are aware of.

\begfigwid 22.3cm
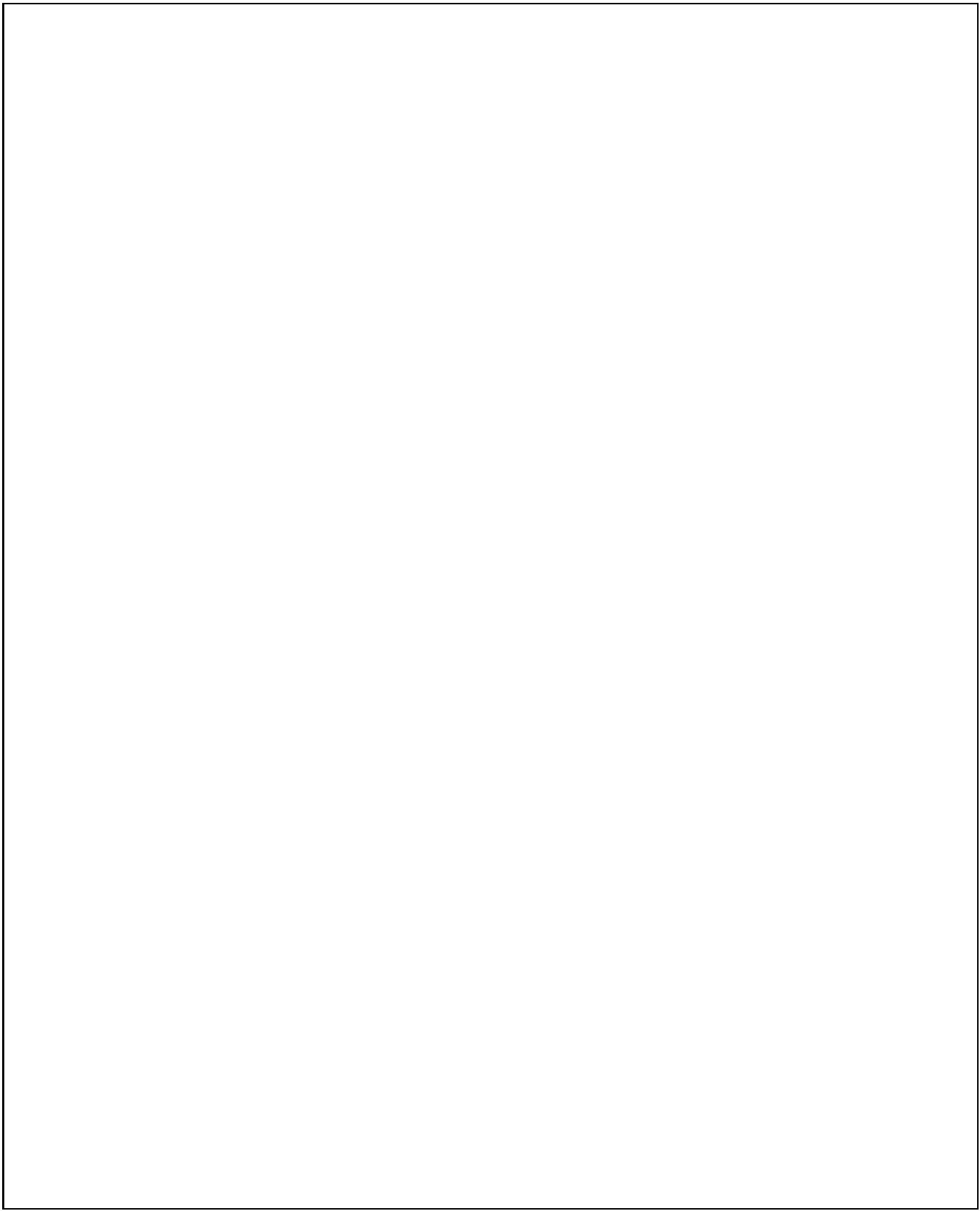
      \figure{1} {Projection of the particles on the (arbitrary)
       x--y--plane at different
       times (from upper left to lower right: $t=0,1.5,2.5,5.0,20.0,200.0$).
       The simulation was performed with $N=32768$ particles and a softening
       length $\varepsilon=0.01$ (model A1).}
\endfig

 Hence direct $N$-body simulations still are
 the appropriate tool to examine cold dissipationless collapse.
 In recent years significant advances in hardware and software for
 direct $N$-body simulations have been made. On one
 hand \tree\ codes (Barnes \& Hut 1986) have been ported
 to parallel computers and coupled with particle based
 hydrodynamic simulations (\tree--SPH, Dav\'e et al. 1997).
 On the other hand completely different approaches like grid--based
 schemes (Udry 1993) or the self-consistent-field method (SCF) by Hernquist \&
 Ostriker (1992) have been succesfully used
 for stellardynamical applications (e.g. in the
 case of cold collapse Hozumi \& Hernquist 1995).
 Although e.g.\ the latter appear to be ideally suited for collisionless
 systems, because they do not use particle-particle forces
 to integrate the orbits, they suffer from their inflexibility
 in adaptation to different applications and resolutions
 (basically for every new application an appropriate set
 of basis functions to evaluate the potential has to
 be searched, and the attempt to increase small-scale
 resolution, e.g. in angular direction, increases the
 CPU time nearly as inhibitively as for direct $N$-body
 simulations, since CPU time goes roughly with $n^2$,
 where $n$ is the number of terms in the series evaluation
 for tangential resolution, much like a force calculation
 scales with $N^2$ where $N$ is the total particle number).

 Thus, even for collisionless systems and applications with
 correspondingly very large particle numbers, direct or
 at least \tree--like $N$-body simulations will remain very 
 useful and important in the foreseeable future because of
 their inherent Lagrangian nature, yielding high resolution at
 high density regions automatically. Moreover, we think
 it is not fully clear what is the physically correct case for
 a collapsing protogalaxy. On one hand, the collapse might
 occur in a system consisting of protogalactic clouds of masses
 as large as say $10^6 {\rm M}_\odot$ (cf. e.g. Theis \& Hensler 1993, 1995);
 in such case a collisionless approach like in the case of SCF codes
 does not represent the physical situation. On the other hand,
 even considering a collapsing system made out of billions of stars
 only, some relaxation effects (not necessarily just two--body relaxation)
 can occur on timescales much shorter than the standard two--body
 relaxation time. It is not clear and deserves further comparison
 between the different approaches whether a mean field based or a direct $N$-body
 method (using large $N$ and very small softening) represent a more
 accurate representation of the real physical system under study.

 For the direct $N$-body simulations
 special purpose computers have been very successfully built
 (Sugimoto et al. 1990, Makino et al. 1997) and applied
 (Makino 1996, Makino \& Ebisuzaki
 1996, Makino 1997). There are different versions
 of such machines e.g. for high precision (collisional stellar
 dynamics, HARP, GRAPE-4) and low precision (collisionless systems,
 GRAPE-3, GRAPE-5). We have
 used the GRAPE 3Af and HARP-2 special purpose
 computers in Kiel and Heidelberg to produce a data base on the influence of
 the softening length and the particle number on the evolution
 of the shape (axis ratios) of stellar systems during cold collapse and
 ROI. Due to the use
 of GRAPE 3Af (4.8 Gflop peak performance) our data could cover
 simultaneously
 small softening and large particle number $N$. The use of HARP-2
 enabled us to provide data without any softening for comparison.
 Small softening should increase two--body relaxation,
 whereas large $N$ decreases it (compared to the dynamical
 timescale). In a real galactic system $\varepsilon$ should
 be vanishing
 for all practical purposes, since the radii of the stars are extremely
 small relative to the mean interparticle distance. On the other hand,
 the particle number is so large as well (e.g. $10^{11}$) that the
 standard two--body relaxation time 
 is large compared to the age of the universe.
 Since a direct model with $N=10^{11}$ and a vanishing softening length
 $\varepsilon$ is impossible
 now and in the near future by a wide margin, generally a system
 with very different $N$ and $\varepsilon$ is used, very often just
 in the vague hope that $\varepsilon$ is large enough to render all
 two--body relaxation effects unimportant during the simulation and to use $N$
 as large as possible with the present computer generation (compare Theis 1998).
 As for the evolution of the shape of triaxial galaxies, emerging after
 cold collapse of an initially extended gas-free stellar system, we will
 show that relaxation effects still play an unwanted role in the simulations.
 As a result we conclude that great care should be taken in interpreting
 results of such $N$-body simulations as models of real galaxies.
  \bigskip

 \titlea{Numerical Results}

 \titleb{The reference model}

  In our reference model (model A1) we started with $N=32768$ particles of total 
mass $M=1$. The particles were distributed according to a Plummer density 
profile 
$$\rho(r) = \rho_0 \cdot (1+(r/r_0)^2)^{-5/2} .\eqno(1)$$
The units are
normalized to a gravitational constant $G=1$, a unit mass and a total
potential energy $U = - 1/2$. For a virialized system the latter choice
guarantees a normalization to the standard N--body units. However, here we
start with a cold system (i.e.\ the total energy $E \approx U$) resulting
in a crossing time $t_{\rm cr} \equiv (GM^{5/2})/\sqrt{2|E|}^3 \approx 1.03$,
for a virial ratio $\eta_{\rm vir} \equiv 2 T / |U| = 0.04$.
($T$ is the total kinetic energy.)
The mass distribution corresponds for a Plummer model to an initial 
scale radius $r_0 = 3 \pi / 16 \approx 0.589$ giving a free--fall time at 
the scale radius of $t_{\rm ff}(r_0) \equiv \sqrt{3\pi / 
(32 G \bar{\rho}(r_0))} \approx 0.84$. $\bar{\rho}(r)$ is the average
mass density inside a sphere of radius $r$.
The initial velocities were chosen according to an isotropic
Gaussian distribution with a velocity dispersion that results in an initial 
virial coefficient $\eta_{\rm vir} = 0.04$. Thus, the collapsing system
is prepared to form a triaxial system via a ROI
(Aguilar \& Merritt 1990).

 The gravitational softening length
was set to $\varepsilon=0.01$ which is a factor of two smaller than the
average inter--particle distance $r_n = r_0 / \cbrt{N}$ at the center and
a factor of 1000 larger than the average $90^\circ$ deflection impact parameter
$p_0 = r_0 / (2 N) \approx 9 \cdot 10^{-6}$. In opposition to select 
$\varepsilon$ large enough to have little undesired two--body relaxation
we here want to examine the different shape evolution of systems with
small and large softening after and during the so-called ``cold collapse''.
It is our aim by this method to study the influence of varying degrees
of two--body relaxation on the results.
The time integration in the simulations was performed with a 
leap--frog scheme using a fixed timestep of $\Delta t = 0.0015$ which is 
less than 0.3 per cent of the central free--fall time. This small timestep
is the price for an acceptable energy error in the numerical simulation 
of the collapse of an initially cold configuration when a small
gravitational softening is applied. The energy conservation is in all 
simulations better than 0.2 per cent, mostly better than 0.05 per 
cent. (Typically the energy error is at maximum in the collapse
phase, whereas lateron the energy fluctuations are an order of magnitude
smaller than the largest deviation.)
The potential was evaluated by a GRAPE 3Af special purpose computer
which applies the standard Plummer softening, i.e.\ the acceleration
${\bf a}_{ij}$ of particle $i$ by particle $j$ is calculated by
 $${\bf a}_{ij} = {Gm_j \over (r_{ij}^2 + \varepsilon^2)^{3/2}}
  {\bf r}_{ij}, \eqno(2) $$
   The temporal evolution of the system starts with a strong collapse. 
After a typical collapse time $t_{\rm ff}(r_0)$ of the system
the central region begins to reexpand whereas the outer layers are still
falling to the center because of their larger free--fall time. Three collapse
times later the inner 75 per cent of the system reach an equilibrium state,
i.e.\ the Lagrangian radii remain almost constant. The half--mass radius 
has decreased by a factor of 1.6 compared to the initial value and the 
inner Lagrangian radii are typically smaller by a factor 3--5. The relatively
small decrease of the half--mass radius is a result of the change in the
shape of the system: During the first collapse phase the system remains 
spherical, which is consistent with the results of Hozumi, Fujiwara 
\& Kan-ya (1996). The shape however changes quickly afterwards, 
when the central region reexpands
(Fig.\ 1). After $t=1.79 t_{\rm ff}(r_0) = 1.5$ a central bar--like structure
can be identified which becomes more and more 
dominant until $t=2.5$. The shape is now
almost fixed in the sense that it does not vary significantly on a crossing 
timescale. However, due to relaxation processes which are discussed below
there is a secular evolution towards a more spherical mass distribution 
(Fig.\ 1). The dynamics of the central part can be followed by the
evolution of the core parameters and the density centre. In Fig.\ 2a
the density weighted core radius $r_{\rm core}$ defined in the same way as 
the density radius in Casertano \& Hut (1985) decreases strongly during 
the collapse phase and reaches an equilibrium value of 0.04 at $t=2.0$.
The corresponding core mass drops from 0.17 for the initial
Plummer model to a final core mass $M_{\rm core} = 0.05$. Once the
equilibrium state is reached, both values
vary only by 10\% during the whole simulated evolution (until $t=200$) which
demonstrates the remarkable constancy of the core properties. The motion
of the density centre will be described later in Sect.\ 3.

\begfig 5.5cm
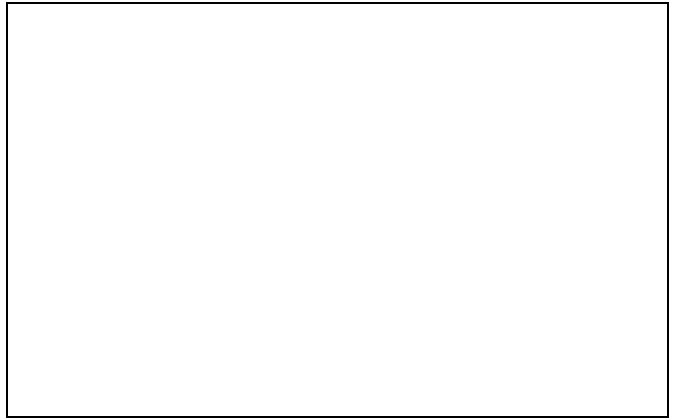
      \figure{2a} {Temporal evolution of the core radius (solid) and the 
      	     core mass (dashed)
             for model A1 (N=32768, $\varepsilon=0.01$).}
\endfig

   The flattening of the mass distribution is calculated by
transformation to the principal axes of the tensor of inertia relative
to the density centre of the system using
different numbers of particles which are sorted by their binding
energy. For 50\% of all
bound particles one finds that the minor to major axis ratio 
$c/a$ begins to decrease at $t=0.8$ and reaches a minimum of 0.36 at 
$t=3.0$ (Fig.\ 2b). Afterwards the axis ratio increases
linearily to 0.52 at $t=20$. The intermediate--to--major axis ratio $b/a$
shows a similar behaviour as $c/a$: With a short delay of $\Delta t = 0.2$
$b/a$ starts to decrease at $t=1.0$ and reaches a minimum of 0.42 at $t=3.0$. 
Lateron, the ratio $c/a$ is on average 0.1 less than $b/a$ which indicates
a slightly triaxial (but near to prolate) system characterized by a 
triaxiality parameter $\tau \equiv (b-c) / (a-c)$ in the range of 0.15 to 0.25.
In the central region, i.e.\ the inner most 10 per cent, the decrease of
the axis ratios begins at the same time as at the half mass radius, but
they are typically larger by 0.1-0.15 (Fig.\ 2b). The flattening of the halo 
region, i.e.\ the 90 per cent most bound particles, is weak and 
strongly delayed according to the enhanced collapse time
in the outer region. Additionally, the isotropic redistribution of particles 
gaining energy in the violent relaxation phase leads to a more spherical
structure of the halo.

\begfig 5.5cm
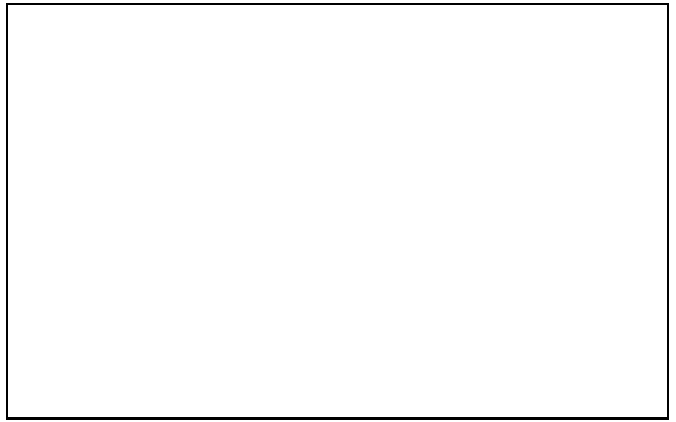
      \figure{2b} {Temporal evolution of the minor--to--major axis ratio
        in the centre (10\% most bounded particles, upper curves after 
        collapse) and at the half--mass radius.
        The softening length was set to $\varepsilon=0.01$ (dashed, model A1) 
        and $\varepsilon=0.1$ (solid, model B1) with $N=32768$. Note here
     and in the following figures that the timescale of the first collapse
     of the central region is constant and is equal to 
     $t_{\rm ff}(r_0) \approx 0.84$ for all models. }
\endfig

\begfig 5.5cm
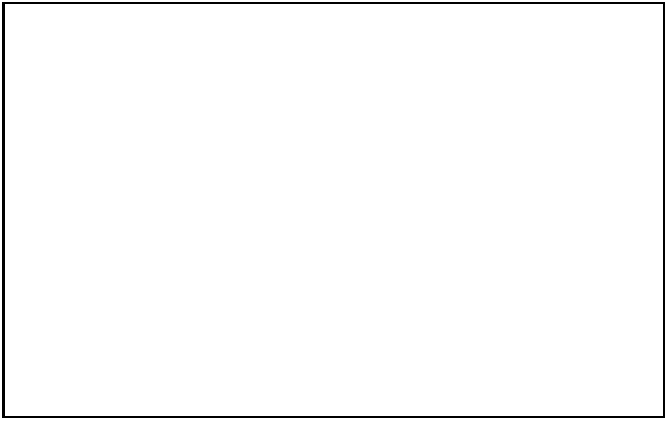
      \figure{2c} {Temporal evolution of the Lagrange radii for a model with 
        $N$=4096, no softening ($+$, HARP) and a simulation with
        $\varepsilon=0.01$ (solid, GRAPE, model D).}
\endfig

\begfig 5.5cm
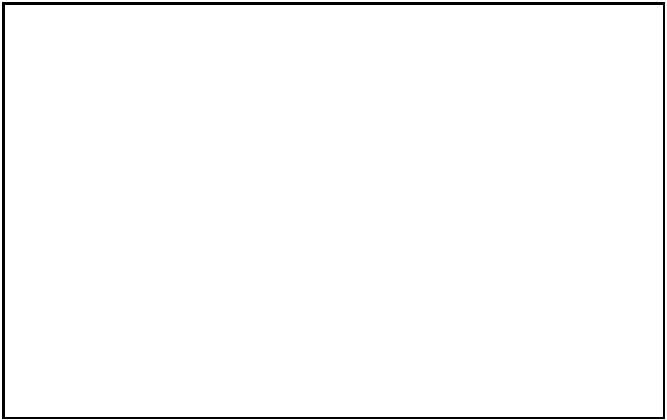
      \figure{2d} {As Fig. 2c, but for $N$=16384 (GRAPE, model F).}
\endfig

\titleb{A parameter study}   

 In order to check the influence of the parameters used in N--body
simulations we selected one reference model (A1) and then
varied its parameters in a sequence of models. Physically we
changed the gravitational softening length $\varepsilon$ (models B1-3 and C),
the total number $N$ of the particles (models D, F),
and $\varepsilon$ and $N$ simultaneously so as to keep the ratio
of $\varepsilon$ to the mean interparticle distance constant (model E).
In order to study the influence of softening on the long--term evolution
of the quasi--equilibrium configuration formed after ROI, two simulations
with different softening, but identical configurations as model F
at $t=20$ were performed (models G, H).
For the two models I and J the initial virial coefficient was
changed from the reference value $\eta=0.04$ to $\eta = 0.2$ in order to
compare our models with the results of Aguilar \& Merritt (1990).
To test the intrinsic errors of our program we varied in models A2 and
B2 the random number set used to initialize the particle configuration,
and in model A3 and B3 a significantly larger (constant) time step was used
(increased by a factor of 10 as compared to the reference model). For model
A3 with the small softening length of $0.01$ it led to untolerable
errors of the energy ''conservation'' ($\sim 90\%$), 
so this model was discarded for the data evaluation (see also Sect.\ 2.2.4). 

For all models with small initial virial coefficient the collapse of
the core proceeded in one free-fall time; Figs.\ 2c,d depict
the evolution of Lagrangian mass shells containing the
indicated fraction of total mass for $N=4096$ and $N=16384$. 
For comparison data of
a run using HARP-2 and a different $N$-body integrator
({\SCS Nbody6++}, Spurzem 1998) without any softening are given.
The excellent agreement of both results tells us that for the reference model
there are no more softening-dependent effects in the simulation during the 
collapse timescale up to the maximum particle number used in our HARP
simulations (16384), hence the softened model is dynamically equivalent to
the ``real'' discrete system. This however, has to be proven again
when increasing $N$ further.

Only after the maximum central density in the collapse is reached
strong deviations of
the shape of the system from spherical symmetry (generally triaxiality)
occur (as can be seen by comparing Figs.\ 2b and 2c,d).
According to Hozumi, Fujiwara \& Kan-ya (1996) this could be
ascribed to the ROI producing strongly anisotropic
velocity dispersions, which then give rise to a non-spherical equilibrium
system. 

Tables 1-3 give an overview for all models presented here,
first the model parameters (Table 1), and thereafter some typical
values describing the flattening (small and intermediate axes relative
to the major axis, and its time derivative) with the timescale $t_{\rm min}$
defined as the time of minimum $c/a$ (which is almost identical to the
time of minimum $b/a$). Note that $t_{\rm min}$ is larger than the 
collapse time of the core.

As a general conclusion from Tables 2 and 3 one could already deduce
that small softening tends to increase the rate of change of $c/a$ after
the system has become flattened.
At a first glance this could be interpreted as a stronger effect
of two--body relaxation present in these systems. However, we will
argue in the following that more caution is necessary in the interpretation
of the data.

\begtabfull
\tabcap{1}{Model parameters for the numerical simulations}
\halign{#\hfil&&\quad#\hfil\cr
\noalign{\hrule\medskip}
model & $N$ & $\varepsilon$ & $\eta_{\rm vir}$ & $\Delta t$ & comment \cr
\noalign{\medskip\hrule\medskip}
A1 & 32768 & 0.01 & 0.04 & 0.0015 & reference model \cr
A2 & 32768 & 0.01 & 0.04 & 0.0015 & different random set \cr 
A3 & 32768 & 0.01 & 0.04 & 0.015  & large energy error \cr   
B1 & 32768 & 0.1  & 0.04 & 0.0015 & \cr
B2 & 32768 & 0.1  & 0.04 & 0.0015 & different random set \cr 
B3 & 32768 & 0.1  & 0.04 & 0.015  & \cr                      
C  & 32768 & 0.02 & 0.04 & 0.0015 & \cr
D &  4096 & 0.01 & 0.04 & 0.0015 & \cr
E &  4096 & 0.02 & 0.04 & 0.0015 & \cr
F & 16384 & 0.01 & 0.04 & 0.0015 & \cr
G & 16384 & 0.02 & 0.04 & 0.0015 & \cr
H & 16384 & 0.1  & 0.04 & 0.0015 & \cr
I & 32768 & 0.01 & 0.2  & 0.0015 & \cr
J &  4096 & 0.01 & 0.2 & 0.0015 & \cr
\noalign{\medskip\hrule}}\endtab

\begtabfull
\tabcap{2}{Flattening parameters of the 50\% most bound particles}
\halign{#\hfil&&\quad#\hfil\cr
\noalign{\hrule\medskip}
model & $t_{\rm min}$ & ${c \over a}_{\rm min}$ & ${b \over a}_{\rm min}$ 
                      & ${c \over a}_{\rm fin}$ & ${b \over a}_{\rm fin}$ \cr
\noalign{\medskip\hrule\medskip}
A1 & 2.9 & 0.36 & 0.41 & 0.48 & 0.57 \cr
A2 & 3.9 & 0.36 & 0.41 & 0.52 & 0.65 \cr
B1 & 6.6 & 0.34 & 0.39 & 0.35 & 0.42 \cr
B2 & 7.1 & 0.37 & 0.39 & 0.38 & 0.42 \cr
B3 & 6.6 & 0.34 & 0.38 & 0.38 & 0.44 \cr
C  & 4.0 & 0.34 & 0.39 & 0.46 & 0.57 \cr
D  & 3.3 & 0.39 & 0.42 & 0.54 & 0.57 \cr
E  & 4.3 & 0.35 & 0.41 & 0.50 & 0.58 \cr
F  & 3.7 & 0.35 & 0.40 & 0.52 & 0.57 \cr
\noalign{\medskip\hrule}}
\smallskip\noindent
The models G and H are omitted because they start with the
configuration of model F at $t=20$.
The dynamically hot models I,J are omitted since they do not
flatten significantly.
Model A3 is omitted because energy conservation is not sufficient.
The final time corresponds here to $t=20$.
\endtab

\begtabfull
\tabcap{3}{Flattening parameters of the 10\% most bound particles}
\halign{#\hfil&&\quad#\hfil\cr
\noalign{\hrule\medskip}
model & $t_{\rm min}$ & ${c \over a}_{\rm min}$ & ${b \over a}_{\rm min}$ 
                      & ${c \over a}_{\rm fin}$ & ${b \over a}_{\rm fin}$ \cr
\noalign{\medskip\hrule\medskip}
A1 & 2.9 & 0.53 & 0.55 & 0.62 & 0.64 \cr
A2 & 4.1 & 0.53 & 0.56 & 0.65 & 0.71 \cr
B1 & 6.8 & 0.78 & 0.80 & 0.77 & 0.80 \cr
B2 & 7.0 & 0.76 & 0.79 & 0.76 & 0.79 \cr
B3 & 6.5 & 0.78 & 0.80 & 0.77 & 0.78 \cr
C  & 4.0 & 0.54 & 0.56 & 0.60 & 0.63 \cr
D  & 2.1 & 0.67 & 0.72 & 0.81 & 0.87 \cr
E  & 4.3 & 0.66 & 0.70 & 0.67 & 0.71 \cr
F  & 2.8 & 0.60 & 0.61 & 0.71 & 0.77 \cr
\noalign{\medskip\hrule}}
\smallskip\noindent
Models G,H,I,J,A3 are omitted for the same reasons as in Table 2.
The final time corresponds here to $t=20$.
\endtab

\titlec{The softening}   

  For the investigation of the effect of softening we increased
$\varepsilon$ by a factor 10 to 0.1 which is a factor of 5 larger than the mean
central interparticle distance (model B1). Fig.\ 2b 
shows that the deviation from
sphericity for half of the particles is now delayed by a factor 2.5--3 
compared to the reference model A1 with a smaller softening. The minimum
axis ratio $c/a$ reaches 0.34 at $t=6.65$ for the 50 per cent most
bound particles which is almost the same minimum value as in model A1.
Until the end of the simulation $c/a$ increases only very slightly to
0.35. 

Here we find two quite different effects of a variation of softening.
First, the build-up of non-sphericity needs more time if softening
is larger. This is valid for both the 10\% and the 50\% most bound
particles. Second the difference between the shape of the
10\% and the 50\% most bound particles is smaller for small softening. Or,
in other words, decreasing the softening makes the 10\% most bound
particles less spherical, but the 50\% most bound particles more
spherical. We interpret such results as follows: first the initial collapse
proceeds to much deeper central potential in the case of small softening.
So the effect of the ROI is stronger, which explains
the difference in $t_{\rm min}$ of the two cases. The term stronger here
means that the timescale for the build-up of
deviations from spherical symmetry 
is smaller, as well as a smaller numerical value of $c/a$
for the 10\% most bound particles. 

\titlec{The number of particles}   

The effect of a variation of the particle number at constant softening
is shown in Fig. 3. While until the time $t_{\rm min}$ differences are very
small, ROI proceeds to stronger perturbations of the spherical shape in
the inner shells (10\%) for larger $N$. 

\begfig 5.5cm
      \figure{3} {Temporal evolution of the minor--to--major axis ratio
        in the centre (10\% most bounded particles) and at the half--mass radius.
        The number of particles has been set to $N=32768$ (solid, model A1) 
        and $N=4096$ (dashed, model D) with a softening length 
        $\varepsilon=0.01$.}
\endfig

\begfig 5.5cm
      \figure{4} {Temporal evolution of the minor--to--major axis ratio
        in the centre (10\% most bounded particles) and at the half--mass radius.
        The number of particles has been set to $N=32768$ (solid, model A1) 
        and $N=4096$ (dashed, model E) with a softening length 
        $\varepsilon=0.01$ (model A1) and $\varepsilon=0.02$ (model E), 
        respectively.}
\endfig

  In the models D and E we decreased the total number of particles to
$N=4096$ and performed the calculation with two different
softening lengths: $\varepsilon=0.01$ (model D) and $\varepsilon=0.02$ 
(model E).
Until $c/a$ reaches its minimum value at $t=3.0$ the temporal evolution of 
the axis ratio inside the 50 per cent region are very similar to 
the reference model A1 with the same softening, but an increased particle 
number of $N=32768$ (Fig.\ 3). We conclude that until $t_{\rm min}$ the ROI
measured at the half--mass radius does not depend strongly on the number 
of particles, but only on the depth of central potential reached. This
supports that the onset of ROI is a collective effect. 

  In model E the softening length is a factor 2 larger than in model A1 and D.
This means that the ratio of the softening length to the average particle 
distance
is the same for the reference model A1 and model E. 
In the central region the larger softening induces a finally increased
$c/a$ of 0.7 (Fig.\ 4). As in the previous case larger $N$ generally
increases the strength of ROI in the inner parts of the system.
However, small differences in the evolution of the 50\% mass shell
can be compensated by our variation of $\varepsilon \sim 1/\cbrt{N}$.
This suggests that gravitational scattering plays a role here. However,
classical two--body relaxation by particle--particle interaction 
(e.g.\ Spitzer \& Hart, 1971) is too slow to explain the observed
compensation. Thus, the scattering of particles with the massive and
much more concentrated core of the system as a whole might cause the 
relaxation here. This point will be discussed in Sect.\ 3 in more
detail. We are grateful to T. Tsuchiya (pers.
communication) to point this out to us as a possible mechanism.

\titlec{The virial coefficient}   

\begfig 5.5cm
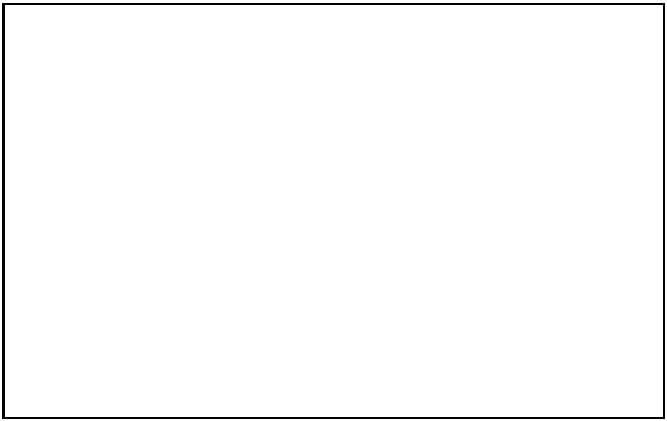
      \figure{5} {Temporal evolution of the minor--to--major ($c/a$) and 
        inter\-mediate--to--major ($b/a$) axis ratio for the
        50\% most bounded particles. The initial virial coefficient has been 
        set to $\eta_{\rm vir} = 0.04$ (solid, model A1) and 
        $\eta_{\rm vir} = 0.2$ (dashed, model I). Both models were calculated
        with $N=32768$ and $\varepsilon=0.01$.}
\endfig

   We studied the influence of the initial distribution in velocity space 
by starting with a dynamically ``warm'' configuration for the spatial
configurations of the models A1 and D, i.e.\ a virial coefficient
$\eta_{\rm vir} = 0.2$ (models I ($N=32768$) and J ($N=4096$)). In
agreement with Aguilar \& Merritt (1990) and Boily et al.\ (1998) 
the memory to the initial conditions is preserved and the system 
remains almost spherical throughout the simulations when starting from 
spherical initial configurations (Fig.\ 5). 
Within statistical errors the models do not
depend on the total number of particles.

\titlec{Miscellaneous}   

  In a set of models we investigated the influence of numerical effects
introduced by the chosen timestep (models A3 and B3) and the random realization
of the initial particle distribution (models A2 and B2). If the timestep is
increased by an order of magnitude (model A3) the energy is only well conserved 
(better than 1.5 per cent) for a model with a large softening of 
$\varepsilon=0.1$ (H), whereas the small softening of the reference model
introduces a very large integration error at the moment of strongest
compression. If we perform the simulation of model A1 with a doubled
timestep, the results are almost identical to the reference model.

   For several models we performed the simulations with other sets of 
initial configurations (e.g.\ models A2 and B2). 
Generally, the deviations of the axis ratios are less than 5 per cent,
until dynamical equilibrium has been established, i.e.\ until $t \approx 5$.
However, the relaxation timescale seems to vary slightly which gives an 
intrinsic error of $c/a$ of 0.04 at the end of most of the simulations 
($t=20$). Thus, our results do not critically depend either on the chosen 
timestep or on the randomly chosen initial particle configuration.

\titleb{The long-term experiments}

\begfig 5.5cm
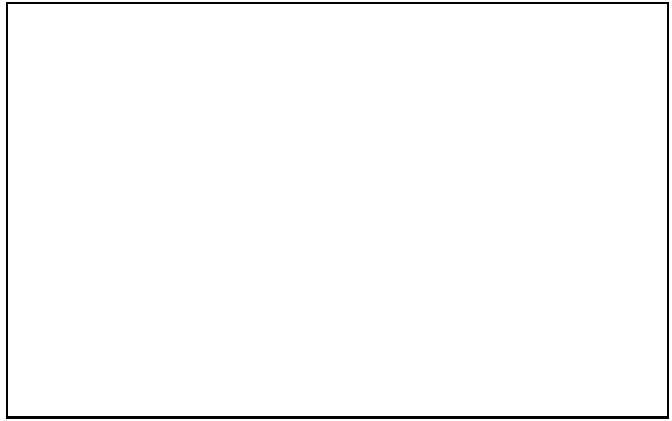
      \figure{6} {Long--term evolution of the intermediate--to--major 
        axis ratio (solid) and the minor--to--major axis ratio (dashed) 
        at the half--mass radius for a simulation with
        $N=16384$ particles. The softening length was set to 
        $\varepsilon=0.01$ (model F).}
\endfig

\begfig 5.5cm
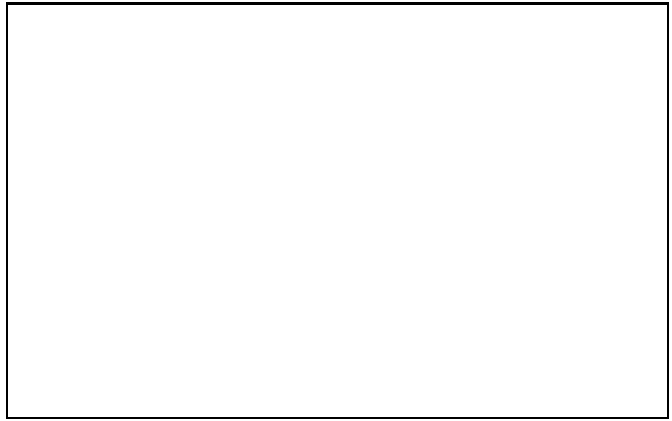
      \figure{7} {Long--term evolution of the minor--to--major axis ratio
        at the half--mass radius for simulations with $N=4096$ (solid, model D)
        and $N=16384$ (dashed, model F) particles. 
        The softening length was set to 
        $\varepsilon=0.01$. The straight lines correspond to a spherization of
        the system in one relaxation time (the latter is already corrected for
        softening).}
\endfig

\begfig 5.5cm
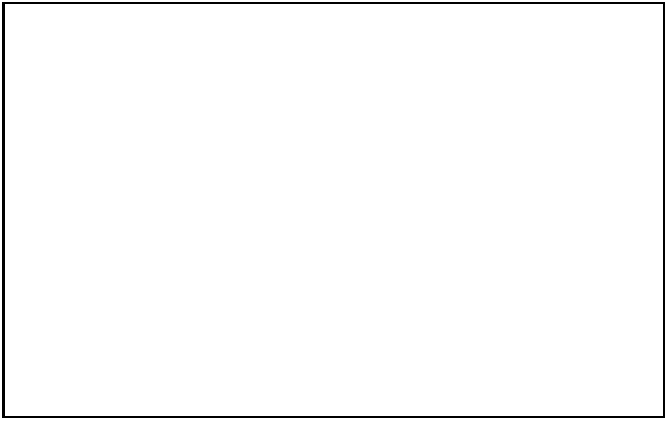
      \figure{8} {Long--term evolution of the (logarithmic) deviation from 
        sphericity at the half--mass radius for simulations with $N=4096$ 
        (solid, model D) and 
        $N=16384$ (dashed, model F) particles. The softening length was set to 
        $\varepsilon=0.01$.}
\endfig

\begfig 5.5cm
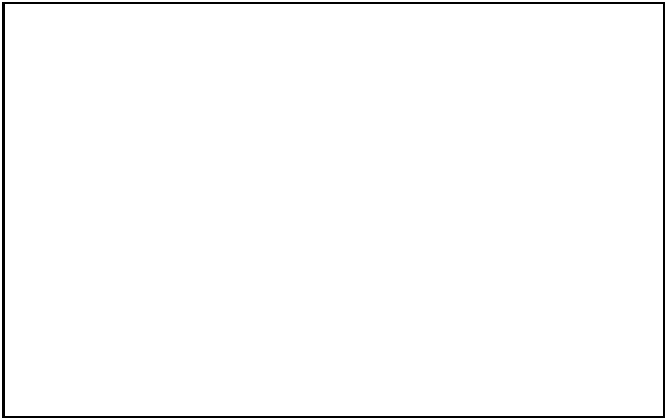
      \figure{9} {Long--term evolution of the (logarithmic) deviation from 
        sphericity at the half--mass radius for simulations with a softening 
        length $\varepsilon=0.01$ (solid, model F) and $\varepsilon=0.02$ 
        (dashed, model G) particles. 
        The simulation was performed with $N=16384$ particles.}
\endfig

\begfig 5.5cm
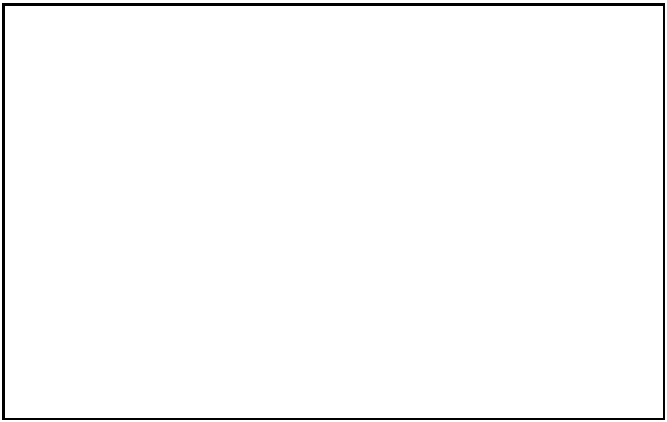
      \figure{10} {Long--term evolution of the (logarithmic) deviation from 
        sphericity at the half--mass radius for simulations with a softening 
        length $\varepsilon=0.1$ (solid, model H) and $\varepsilon=0.02$ 
        (dashed, model G) particles. 
        The simulations were performed with $N=16384$ particles.}
\endfig

In some cases we followed the evolution of the system for a much
longer time, until it practically relaxed again to a spherically
symmetric configuration (models A1, C, D, E, F, G and H). 
The evolution of shape shows three different phases (e.g.\ Figs.\ 6 and 11): 
first after the collapse the ROI leads to a triaxial configuration. 
In the second stage a
fast readjustment of shape occurs during about 20 crossing times.
Thereafter a steady relaxation follows until the system becomes axisymmetric
($b/a\approx1$, but still $c/a \ne 1$) and finally spherical (Fig.\ 6). 
Lateron the system does not change its shape (Fig.\ 7).

The timescale for spherization is given remarkably exact by the 
two--body relaxation time $t_{\rm rlx}$ 
of the N--body system. This is demonstrated in Fig.\ 7
by the temporal evolution of the minor--to--major axis ratio for two
simulations with $N=4096$ and $N=16384$ and a softening length of 
$\varepsilon=0.01$. The straight lines are no fits, but the slopes one gets
by the assumption that spherization acts on a relaxation timescale.
It should be noted that the good agreement can only be achieved when 
the Keplerian relaxation time is corrected for softening (for details
see Theis (1998)).
This gives a correction, i.e.\ an increase of the relaxation times,
by approximately a factor of $3$ for $N=16384$.

   The deviation from sphericity, $\delta(t) \equiv 1 - c/a(t)$, can be
fitted by $\delta(t) \approx \delta(0) \exp(-t/t_{\rm rlx})$ as the extended
linear regime in Fig.\ 8 shows. Both, the dependence of the axis ratio
evolution on the total number of particles (Fig.\ 8) and the 
softening length (Fig.\ 9 and 10) are in agreement with a spherization 
within a relaxation timescale of the softened two--body potential.
The slow relaxation in model H ($\varepsilon = 0.1$) is due to the
small ratio of maximum impact parameter to softening length
($\varepsilon$ is twice the core radius!) which gives a strongly enhanced
relaxation time.
 
Shortly after the initial collapse at $t_{\rm min}$, however, the
``relaxation'' of shape of our models is much faster than one could
expect from the two-body relaxation argument. This can be seen
clearly in Fig.\ 11, which shows the long--term evolution of our
reference model A1 ($N=32768, \varepsilon=0.01$). We denote this phase
of fast shape change by the term ``violent shape relaxation''.
It is charcterized by a substantially larger growth rate of
the axis ratio as compared with expectations from two-body
relaxation. Violent shape relaxation sets in after the ROI has established 
a triaxial configuration. Core radius and core mass do not vary strongly
during violent shape relaxation (see Fig.\ 2a).
Independent of taking 10\% or 50\% of the particles,
the axis ratio growth rate during violent shape relaxation
is constant for about 10--20 crossing times; thereafter
a smaller growth rate obtained by two-body relaxation is following.
Note that all variations of shape seen in Figs. 2-5 and discussed in
subsection 2.2 are still in the violent phase and do not represent the
long-term shape changes from two-body relaxation, which can be seen in
the Figs. 6-11 for the long-term evolution.

\begfig 5.5cm
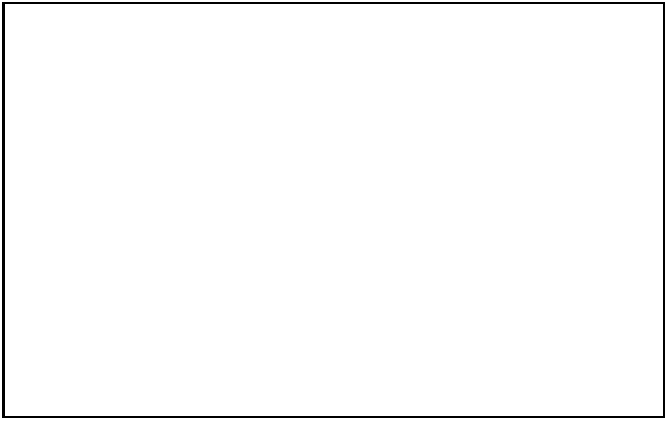
      \figure{11} {Long--term evolution of the minor--to--major axis ratio
          in the centre (10\% most bounded particles, dashed)
          and at the half--mass radius (solid) for model A1
          (N=32768, $\varepsilon=0.01$).}
\endfig
 
\begfig 5.5cm
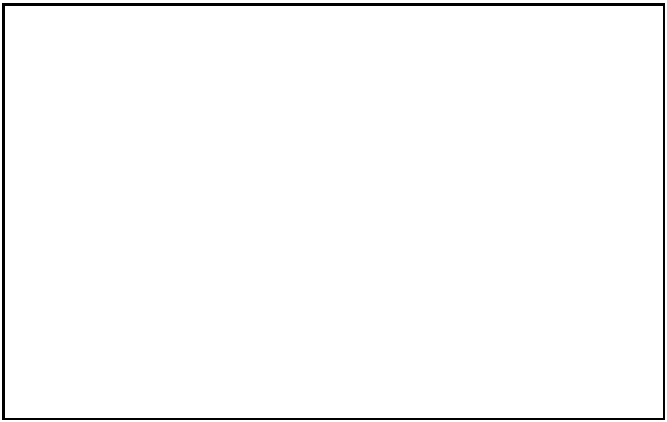
      \figure{12} {Temporal evolution of the distance of the density centre
         from the centre of mass (solid) in correlation with Lagrangian radii
         (0.1\%, 20\%, 50\%, dashed) for model A1 (N=32768, $\varepsilon=0.01$).}
\endfig

\begfig 5.5cm
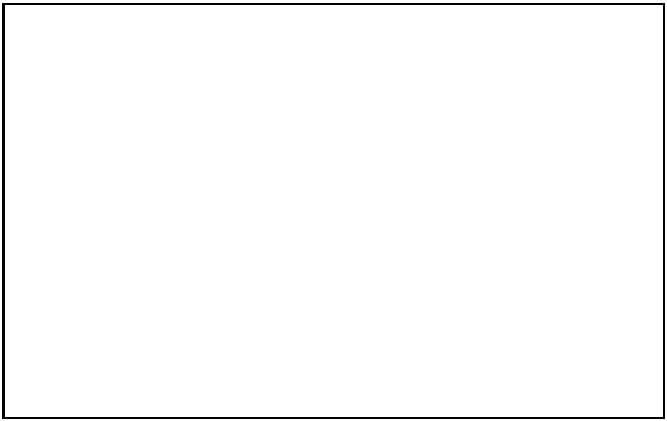
      \figure{13} {Long--term evolution of the minor--to--major axis ratio
        at the half--mass radius for simulations with a softening length
        $\varepsilon=0.01$ (solid, model A1) and $\varepsilon=0.02$ 
        (dashed, model C). 
        The simulations were performed with $N=32768$ particles.}
\endfig

 \titlea{Discussion and Summary}

 We performed a set of $N$--body simulations for the collapse
of a dynamically cold Plummer sphere and studied the occurrence
of radial orbit instability (ROI) and the evolution of shape. 
Especially, we focussed on the
influence of the standard N--body parameters like the total
number of particles or the softening length. Our simulations
showed that the growth rate of the ROI is strongly 
changed by the softening at constant particle number and almost 
unaffected by a variation of $N$ at constant softening. For our standard
models with $\varepsilon =0.01$ and $N=16384$ ($N=4096$),
however, we could show by comparison with high-accuracy zero softening
models, that no more artificial softening-dependent effects are present.
But our results show that for a given particle number inappropriate
softening could yield wrong results by changing the collective effects
as well as the undesired two-body relaxation. Such findings are consistent
with recent arguments given by Steinmetz \& White (1997), Moore et al. (1997),
and Fukushige \& Makino (1997),
in the context of $N$-body simulations for cosmological structure formation.

Generally, according to our simulations, the cold collapse with ROI
and the subsequent evolution could be subdivided into three phases. First
there is an initial strong collapse and build-up of triaxiality via
ROI, characterized by a time $t_{\rm min}$ at
which a minimum value of the principal axis ratios $c/a$ and $b/a$ is
obtained. Here our results agree with the findings of Hozumi, Fujiwara
\& Kan-ya (1996) insofar, that the system remains spherical until
maximum collapse $t_{\rm coll} < t_{\rm min}$; 
it is only at that stage that strong anisotropies
of the velocity dispersions create a triaxial configuration. 
This is also consistent with arguments by Bettwieser \& Spurzem (1986) and
Hensler et al. (1995), who state that homologously collapsing systems have
to stay isotropic.

Second, a phase of rapid, violent shape relaxation occurs. It is much
faster than by any estimate due to two-body relaxation.
The mechanism of such violent shape
relaxation is still unclear. We observe, however, 
strong fluctuations of
the density centre of the system just during that
phase, which are strongly correlated with
oscillations of
the Lagrangian radii of up to 20\% of the total mass 
(see Fig.\ 12, here Lagrangian radii have not been measured with respect 
to the density centre but with respect to the centre of mass of the system). 
Even the half--mass radius varies due to these quasi--oscillations of the core
whereas the 0.1\% Lagrange radius
is completely inside the core and, thus, not affected by the core's motion.
We conclude that
the rapid initial collapse induces strong motions of the density centre,
which decay on a timescale of some ten crossing times.
However, we saw that the core is already formed after two crossing times
and does not change any more. This means, that a core fixed in size
moves in the central region, by this inducing strong potential
fluctuations which give rise to a relaxation which is added to the
unavoidable two--body relaxation. When the core settles down
these large--scale potential fluctuations disappear and the additional
relaxation does not operate any more by this defining the last stage
of shape evolution. A comparison of two long--term simulations with 
different softening lengths strengthens that point, insofar as it
seems that two--body relaxation does not change the violent 
shape relaxation (Fig.\ 13): Whereas a difference in the relaxation rate 
in the last stage after $t=20$ is obvious, the shape evolution is qualitatively
almost unaffected during the violent shape relaxation stage. However, the 
relatively short timescale of violent shape relaxation and the scatter
in the measured $c/a$--values makes a quantitative comparison impossible.
It is interesting to compare with the similar
problem of core oscillations in quasi-statically evolving spherical
star cluster undergoing core collapse (Makino \& Sugimoto 1987, Heggie,
Inagaki \& McMillan 1994, Spurzem \& Aarseth 1996), which seem to
occur on a similar timescale.

Finally, a third phase is observed, which is characterized by a secular drift 
of the system via axisymmetry towards spherical symmetry. It occurs on a 
time scale consistent with that of two--body relaxation,
provided the proper softening is taken into account (Theis 1998).

 \acknow{This work was supported by the DFG grants
 Sp 345/5-1,2,3 and 446 JAP-113/18/0 (for travel). 
 R.Sp.\ and Ch.Th.\ wish to thank
 D. Sugimoto, J. Makino, M. Taiji, and all other
 colleagues from Dept.\ of Arts and Sciences, Komaba
 Campus, Univ.\ of Tokyo, for their continuous support
 in all GRAPE-related problems and their warm hospitality
 during our visits in Tokyo. Numerical computations were
 carried out on GRAPE 3Af (Kiel) and HARP-2 (Heidelberg)
 special purpose computers and partly at HLRZ J"ulich.
 Last, but not least the authors want acknowledge the
 valuable comments of the referee, Roger Fux. }
 \begref{References}
 \ref
 \usrref{\ApJ}{Aguilar L.A., Merritt D.}{1990}{354}{33}
 \usrref{\MN}{van Albada, T.S.}{1982}{201}{939}
 \usrref{\SA}{Antonov V.A.}{1961}{4}{859}
 \ref Antonov V.A.: 1962, {\sl Vest. Leningrad Gos. Univ.} {\bf 7},
 135.
 \usrref{\Nat}{Barnes, J., Hut, P.}{1986}{324}{446}
 \ref Binney J., Tremaine, S.: {\it Galactic Dynamics}, Princeton
    Univ.\ Press (1987)
 \usrref{\AA}{Bettwieser E., Spurzem R.}{1986}{161}{102}
 \usrref{\MN}{Boily C.M., Clarke C.J., Murray S.D.}{1998}{}{in press}
 \usrref{\MN}{Burkert A.}{1990}{247}{152}
 \usrref{\ApJ}{Cannizzo J.K., Hollister T.C.}{1992}{400}{58}
 \usrref{\ApJ}{Casertano S., Hut P.}{1985}{298}{80}
 \usrref{\MN}{Cuperman S., Goldstein S., Lecar M.}{1969}{146}{161}
 \usrref{\NewA}{Dav\'e R., Dubinski J., Hernquist L.}{1997}{2}{277}
 \usrref{\ApJ}{Fukushige T., Makino J.}{1997}{477}{L9}
 \usrref{\MN}{Heggie D.C., Inagaki S., McMillan S.L.W.}{1994}{271}{706}
 \usrref{\AA}{Hensler G., Spurzem R., Burkert A., Trassl E.}{1995}{303}{299}
 \usrref{\ApJ}{Hernquist L., Ostriker J.P.}{1992}{386}{375}
 \usrref{\PASJ}{Hozumi S., Fujiwara T., Kan-ya Y.}{1996}{48}{503}
 \usrref{\ApJ}{Hozumi S., Hernquist L.}{1995}{440}{60}
 \usrref{\ApJ}{Ipser J.R., Thorne K.S.}{1968}{154}{251}
 \usrref{\ApJ}{Kulsrud R.M., Mark J.W-K.}{1970}{160}{471}
 \ref Lynden-Bell D.: 1964, in {\sl IAU symp. No. 25}, p. 78.
 \usrref{\MN}{Lynden-Bell D.}{1967}{136}{101}
 \usrref{\ApJ}{Makino J.}{1996}{471}{796}
 \usrref{\ApJ}{Makino J.}{1997}{478}{58}
 \usrref{\ApJ}{Makino J., Ebisuzaki T.}{1996}{465}{527}
 \usrref{\PASJ}{Makino J., Sugimoto D.}{1987}{39}{589}
 \usrref{\ApJ}{Makino J., Taiji M., Ebisuzaki T., Sugimoto D.}{1997}{480}{432}
 \usrref{\MN}{Merritt D., Aguilar L.A.}{1985}{217}{787}
 \usrref{\MN}{Min K.W., Choi C.S.}{1989}{238}{253}
 \usrref{\ApJ}{Moore B., Governato F., Quinn T., Stadel J., Lake G. }{1997}{
 }{preprint~}{astro-ph 9709051}
 \usrref{\SA}{Polyachenko V.L., Shukhman I.G.}{1981}{25}{533}
 \usrref{\ApJ}{Spitzer L., Hart M.H.}{1971}{164}{399}
 \usrref{\MN}{Spurzem R.}{1998}{}{subm.}
 \usrref{\MN}{Spurzem R., Aarseth S.J.}{1996}{282}{19}
 \usrref{\MN}{Steinmetz M., White S.D.M.}{1997}{288}{545}
 \usrref{\Nat}{Sugimoto D., Chikada Y., Makino J., Ito T.,
    Ebisuzaki T., Umemura M.}{1990}{345}{33}
 \usrref{\AA}{Theis Ch., Hensler G.}{1993}{280}{85} 
 \ref Theis Ch., Hensler G.: 1995, in {\sl Galaxies in the Young Universe},
      Hippelein H. (ed.), Ringberg, p.\ 201
 \usrref{\AA}{Theis Ch.}{1998}{330}{1180}
 \usrref{\AA}{Udry S.}{1993}{268}{35}
 \usrref{\MN}{Wiechen H., Ziegler H.J., Schindler K.}{1988}{232}{623}
 \usrref{\MN}{Ziegler H.J., Wiechen H.}{1989}{238}{1261}
 \usrref{\ApJ}{Ziegler H.J., Wiechen H.}{1990}{362}{595}
 \endref
 \bye